\documentclass[fleqn,]{wlscirep}
\usepackage[utf8]{inputenc}
\usepackage{algorithm}
\usepackage{algpseudocode}
\usepackage[T1]{fontenc}
\usepackage{bm}
\usepackage{float}
\usepackage{multirow}
\usepackage{wrapfig} 
\usepackage{makecell} 
\usepackage{quantikz}
\usepackage{subcaption}
\usepackage{tabularx}
\usepackage{graphicx}
\usepackage{pdflscape}
\usepackage{caption}
\usepackage{xcolor}
\usepackage{longtable}
\usepackage{siunitx}
\usepackage[normalem]{ulem}

\title{Application-Oriented Evaluation of CPU/GPU Quantum Simulators: From Foundational Circuits to Quantum Chemistry}

\author[1]{\small Anurag K. S. V.}
\author[1]{\small Ashish Kumar Patra}
\author[1]{\small Afra Navas}
\author[1, 2]{\small Ruchika Bhat}
\author[1]{\small Sai Shankar P.}
\author[3]{\small Rahul Maitra}
\author[4]{\small Manas Mukherjee}
\author[, 1]{\small Jaiganesh G.\thanks{(Corresponding Author) email: jaiganesh@qclairvoyance.in, drjaiganesh15@gmail.com}}

\affil[1]{\small Qclairvoyance Quantum Labs, Secunderabad, TG 500094, India.}
\affil[2]{\small The University of Arizona, Tucson, AZ 85721, USA.}
\affil[3]{\small Department of Chemistry, Indian Institute of Technology Bombay, Mumbai, MH 400076, India.}
\affil[4]{\small Centre for Quantum Technologies, National University of Singapore, Singapore 117543, Singapore.}

\begin{abstract}
Quantum computing simulators underpin the development and prototyping of quantum algorithms, yet the practical choice between competing software stacks and their CPU and GPU backends is rarely guided by application-oriented evidence. We present a systematic, quantum hardware-agnostic evaluation of two widely used stacks, CUDA-Q and Qiskit, selected as the simulators closest to production execution paths, across CPU and GPU backends, and across single- and double-precision arithmetic. The evaluation spans foundational utility circuits (GHZ-state preparation and QFT/QFT\textsuperscript{$\dagger$} mirror circuits) through applied quantum chemistry (Variational Quantum Eigensolver (VQE) energy estimation with standard UCCSD and a depth-reduced LCNot-UCCSD ansatz, and direct sampling on hydrogen chains up to 32 qubits). The results yield concrete selection guidance: GPU-accelerated double precision delivers consistent speedups, typically $2\times$ to $10\times$ for tractable application workloads and substantially larger at the widest circuits up to $100\times$ or more, and governs the maximum simulable system size under fixed compute memory, while CPU statevector backends and Qiskit Aer remain competitive, and at times preferable, at low qubit counts and wide but shallow circuits. We demonstrate the practical payoff of this guidance by extending an in-house developed QSCI(LCNot-UCCSD)-RBM electronic-structure workflow, previously constrained to a 16-qubit active space on a CPU, to a 32-qubit (10 electrons, 16 orbitals) active space with GPU acceleration, completing a 20-point potential-energy-surface scan of N\textsubscript{2} in the cc-pVDZ basis. This end-to-end evaluation establishes that the principled simulator selection translates directly into expanded scientific reach and faster algorithmic prototyping.
\end{abstract}

\begin{document}

\flushbottom
\maketitle

\thispagestyle{empty}


\newpage

\section{Introduction}
\label{sec:introduction}

The classical simulation of quantum computers is the scaffolding on which the entire prototyping pipeline of quantum-algorithm research is built. Long before a candidate algorithm reaches quantum hardware, it is conceived, debugged, and validated on a classical simulator, where the full quantum state is accessible, measurement statistics are exact, and results are uncontaminated by any physical device noise. This role is structural where the original case for quantum computation was itself framed as the difficulty of reproducing quantum dynamics on classical machines~\cite{Feynman1982, Lloyd1996, Abrams1999}, and the same exponential growth of the Hilbert space that motivates quantum hardware, makes faithful simulation a demanding, resource-bound task~\cite{Xu2025, Cicero2025}. In the noisy intermediate-scale quantum (NISQ) regime~\cite{Preskill2018}, where physical qubit counts and coherence times remain limited, simulators are not merely a convenience but one of the primary venues in which most algorithmic development actually takes place.

Nowhere is this dependence sharper than in quantum computational chemistry, which remains one of the most actively pursued near-term applications of quantum computers~\cite{McArdle2020, Zhang2025}. The variational quantum eigensolver (VQE) and the broader family of variational hybrid algorithms estimate molecular ground-state energies by repeatedly preparing and measuring a parameterised trial state under classical optimisation~\cite{Peruzzo2014, McClean2016, Kandala2017, Tilly2022, Fedorov2022}. Chemically motivated ans\"atze such as unitary coupled cluster with single and double excitations (UCCSD) provide systematically improvable accuracy~\cite{Barkoutsos2018, Romero_2019}, but at a circuit depth that grows rapidly with the active space, which has driven a sustained effort toward Hamiltonian-, depth-, and gate-reduced constructions~\cite{Grimsley2019, Ramoa2025, AKSV2026}. Because a single potential-energy-surface (PES) scan demands many such evaluations across geometries, and because trainability pathologies are diagnosed most cleanly in a noiseless setting~\cite{McClean2018, Wang2021}, the throughput and the maximum tractable system size of the underlying simulator directly govern what chemistry can be prototyped at all. The practical question for a chemistry-oriented practitioner is therefore not whether to simulate, but which software stack, hardware backend, and numerical precision to commit to for a given workload.

The simulator ecosystem that may answer this question has matured into two broad algorithmic families and two dominant hardware-acceleration strategies. State-vector simulation stores the full amplitude vector and is the workhorse for the dense, highly entangled circuits typical of chemistry, while tensor-network methods exploit structure to reach larger but more restricted circuits~\cite{Bayraktar2023}. Acceleration proceeds either by multi-threaded and distributed execution on CPUs~\cite{Tabuchi2023} or by offloading the amplitude updates to GPUs, where the regular, memory-bound structure of the state-vector kernel maps well onto massively parallel hardware~\cite{Bayraktar2023, Doi2023, Jiang2025, Lu2025, Efthymiou2022}. Two such prominent stacks that sit close to production execution paths and anchor this study include Qiskit, whose Aer backend offers tuned CPU state-vector and shot-based sampling engines~\cite{Doi2023}, and CUDA-Q, which exposes the cuQuantum libraries through a unified programming model spanning CPU and GPU targets and single- and double-precision arithmetic~\cite{Bayraktar2023, Kulkarni2026}. It can also be noted that the precision axis is consequential in its own right: reduced-precision and tensor-float formats can yield large throughput gains for amplitude-dominated workloads~\cite{Modani2022, Faj2023}, but trade against the numerical fidelity required to satisfy the chemical accuracy threshold.

Recent benchmarking efforts have begun to chart this landscape, where GPU-acceleration case studies quantify the speedups attainable on state-vector simulation for individual stacks and kernels~\cite{Faj2023, Modani2022, More2025}; quantum-volume and state-vector package comparisons rank simulators on standardized circuit families~\cite{Boehme2024, Amit2024}; software-engineering benchmarks characterise the cost of circuit construction, manipulation, and compilation rather than execution~\cite{Nation2025}; low-level suites such as QASMBench supply standardised NISQ workloads for cross-platform comparison~\cite{Li2023}; and review articles survey acceleration opportunities across CPU, GPU, and FPGA hardware~\cite{Cicero2025, Xu2025}. A complementary line of work has interrogated the correctness and reliability of these tools through statistical assertions for CUDA-Q programs~\cite{Li2025}, empirical studies of simulator bugs~\cite{Upadhyay2026}, and formal verification of the open-source ecosystem~\cite{Blain2026}. Collectively, these works establish that backend choice matters; what they do not yet provide is a single, coherent basis for choosing among them for quantum chemistry application workloads. The decision of which simulator to adopt is thus frequently made on heuristic or institutional grounds rather than on evidence.

This paper addresses this gap with a systematic, quantum hardware-agnostic evaluation of Qiskit and CUDA-Q that spans both CPU and GPU backends and both single- and double-precision arithmetic, and that is carried end to end from foundational utility circuits through to a production molecular-simulation workflow. The evaluation is organised as a chain of six experiments in which the outcome of each motivates the next, narrowing the question by successive elimination. It begins with GHZ-state preparation and a QFT/QFT$^{\dagger}$ mirror circuit that probes entangling-gate throughput, width scaling, and round-trip numerical fidelity; proceeds to a minimal VQE workload on H\textsubscript{2} that fixes the single- versus double-precision accuracy floor and contrasts the CUDA-Q and CUDA-QX programming routes; advances to an applied comparison of standard UCCSD against the linear-CNOT LCNot-UCCSD ansatz~\cite{Magoulas2023, Magoulas2023b, KSV2026} across five molecules in STO-3G, and to direct computational-basis sampling on hydrogen chains H\textsubscript{2}--H\textsubscript{16} (4 to 32 qubits); and culminates in an application workload that converts the accumulated selection guidance into expanded scientific reach. The elimination chain converges on CUDA-Q's \texttt{nvidia:fp64} state-vector backend paired with LCNot-UCCSD as the justified configuration for the terminal application. Concretely, we extend an in-house developed QSCI(LCNot-UCCSD)--RBM electronic-structure workflow~\cite{Patra2026mlcsg, KSV2026FX700}, quantum-selected configuration interaction (QSCI)~\cite{Kanno2023, Nakagawa2024} coupled to a restricted Boltzmann machine (RBM) generator~\cite{Nomura2017, Melko2019RBMQuantumPhysics}, from a $(8e,8o)$/16-qubit active space previously computed on CPU using the Fujitsu QARP stack~\cite{KSV2026FX700} to a $(10e,16o)$ active space (16 spatial orbitals, 10 active electrons, the two nitrogen $1s$ cores frozen; 32 qubits) in the cc-pVDZ basis. This doubles the correlation window from 8 to 16 spatial orbitals, corresponding to 16 to 32 qubits under GPU acceleration, and completes a 20-point PES scan of N\textsubscript{2} that stays within the chemical accuracy threshold of the active-space CASCI$(10e,16o)$ reference across the entire curve, including the strongly correlated triple-bond dissociation region, whereas a sample-based quantum diagonalisation (SQD) baseline~\cite{Robledo2025} recovers essentially no correlation beyond the mean field due to lack of any error for stochastic configuraiton recovery. The principal contributions are:

\begin{enumerate}
  \item A controlled, end-to-end benchmarking protocol for Qiskit and CUDA-Q that jointly varies stack, CPU/GPU backend, and floating-point precision, applied
  uniformly from utility circuits to molecular workloads and reported with resource-estimation and hardware-aware scaling metrics rather than isolated timings.

  \item A foundational benchmark layer (Experiments~1--3) establishing baseline framework behaviour on GHZ-state preparation, a QFT/QFT$^{\dagger}$ mirror circuit, and a minimal VQE workload on H\textsubscript{2}, characterising per-backend width scaling, shot-density sensitivity, and the single- versus double-precision accuracy floor, where single precision both lowers the qubit ceiling by one and fails to reach the chemical accuracy for energy convergence.

  \item A quantum chemistry algorithms benchmark layer (Experiments~4--5) comparing UCCSD against LCNot-UCCSD under VQE energy estimation across five molecules, and evaluating direct sampling on hydrogen chains up to 32 qubits, linking the ansatz's linear-CNOT construction, which reduces gate counts by up to $14\times$ relative to UCCSD, to end-to-end runtime and to the compute memory-bounded maximum simulable width.

  \item Concrete, evidence-based selection guidance rather than a single ranking: GPU acceleration amortises its fixed overhead only above roughly 18--26 qubits
  (workload-dependent), below which CPU state-vector backends remain competitive, Qiskit Aer is strongest at shallow circuits across all shot densities, and the Qiskit
  state-vector sampler is fastest at the smallest widths, while the GPU wins its advantage ranges from about $2\times$ on the gate-heavy 32-qubit application sampling workload, $5\times$ on LCNot-UCCSD VQE, up to $126\times$ on 31-qubit GHZ preparation, and up to $110\times$ for  QFT/QFT$^{\dagger}$ mirror circuits.

  \item A quantum chemistry application demonstration (Experiment~6) that operationalises this guidance by doubling the active-space width of the QSCI(LCNot-UCCSD)--RBM workflow from 16 to 32 qubits under GPU acceleration and completing a 20-point N\textsubscript{2} PES scan within chemical accuracy of CASCI$(10e,16o)$ on a single workstation GPU in double precision, a surface out of reach of the earlier 16-qubit CPU workflow.
\end{enumerate}

The remainder of this paper is organised as follows. Section~\ref{sec:methodology} details the methodology, including the two software stacks, the per-phase hardware configurations, the six experimental designs, and the resource-estimation and precision metrics used throughout. Section~\ref{sec:results} presents and discusses the empirical results across the foundational and application tiers. Section~\ref{sec:conclusion} concludes and outlines directions for extending the work.
\section{Methodology}
\label{sec:methodology}

The comparison of Qiskit with the Aer statevector backend on the CPU host and CUDA-Q with its GPU statevector backend is organised as a six-experiment motivational chain across three phases. Phase~1 establishes framework behaviour on structurally transparent utility circuits (GHZ preparation, a QFT--QFT$^{\dagger}$ mirror, and VQE on H$_2$). Phase~2 moves to chemically meaningful workloads (a UCCSD versus LCNot-UCCSD variational comparison across five molecules, and LCNot-UCCSD quantum sampling on hydrogen chains up to 32 qubits). Phase~3 is the culminating application: a nitrogen-molecule potential-energy-surface (PES) scan driven by an in-house-developed QSCI(LCNot-UCCSD)--RBM workflow, extended from 16 to 32 qubits with GPU acceleration. Each experiment is designed so that its outcome motivates the next, progressively narrowing the question from ``which backend is faster'' to ``which backend, at which arithmetic precision, sustains an abstracted fault-tolerant-regime chemistry workload at the 32-qubit scale.'' Throughout, the initial axes of interest are wall-clock timing, compute memory footprint, and single- versus double-precision (fp32/fp64) arithmetic, followed by quantum resource estimation (qubit width, gate counts) and the accuracy of the obtained results.

\subsection{Common Software Environment}
\label{subsec:software}

A single, version-pinned software environment was used for every experiment to remove toolchain drift as a confounding variable. Circuit construction, transpilation, and CPU simulation used Qiskit~\cite{Ali2024} (\texttt{qiskit 1.4.2}) with the Aer statevector simulator (\texttt{qiskit-aer 0.17.0}); SQD and VQE relied on \texttt{qiskit-addon-sqd 0.10.0}~\cite{QiskitAddonSQD2025} and \texttt{qiskit-nature 0.7.2}. GPU simulation used the CUDA-Q platform \texttt{cuda-quantum-cu12 0.12.0} together with \texttt{cudaq-solvers 0.4.0} for its chemistry state-preparation and VQE primitives. Electronic-structure preprocessing (RHF, MP2, CCSD, and FCI reference values) was performed with PySCF~\cite{Sun2015, Sun2018, Sun2020} (\texttt{pyscf 2.10.0}); molecular Hamiltonians for the CUDA-Q path were built through OpenFermion (\texttt{openfermion 1.7.1}) and \texttt{openfermionpyscf 0.5}. The restricted Boltzmann machine used in the N$_2$ scan was implemented in PyTorch (\texttt{pytorch 2.11.0}). Numerical support was provided by NumPy~\cite{Harris2020} (\texttt{1.26.4}), SciPy~\cite{Virtanen2020} (\texttt{1.15.3}), and JAX~\cite{Jax2018} (\texttt{0.5.3}); GPU utilisation and memory were monitored with \texttt{pynvml 12.0.0}, and results were aggregated with pandas~\cite{Pandas2020} (\texttt{2.2.3}) and written incrementally to CSV to guard against interruption.

\subsection{Physical Hardware}
\label{subsec:hardware}

Three physical platforms were used, mapped to the three experimental phases as summarised in Table~\ref{tab:hardware}. Because the study contrasts a CPU-resident framework (Qiskit/Aer) with a GPU-resident one (CUDA-Q), each platform is characterised by both its host CPU (which carries the Qiskit workload and all classical pre- and post-processing) and its accelerator (which carries the CUDA-Q workload). Keeping the CPU--GPU pairing fixed within a phase ensures that any measured CPU-versus-GPU difference is attributable to the frameworks and their precision settings rather than to a change of machine mid-experiment.

\begin{table*}[htbp]
\centering
\caption{Physical hardware used in this work, mapped to the six-experiment chain. The Qiskit/Aer workload and all classical electronic-structure pre- and post-processing execute on the host CPU; the CUDA-Q workload executes on the accelerator. Within each phase, the CPU--GPU pairing is held fixed.}
\label{tab:hardware}
\small
\begin{tabular}{@{}p{2.7cm} p{3.1cm} p{3.6cm} p{4.6cm}@{}}
\toprule
\textbf{Phase / Experiments} & \textbf{Platform} & \textbf{Host CPU (Qiskit/Aer)} & \textbf{Accelerator (CUDA-Q)} \\
\midrule
Phase~1 \newline (Exp.~1--3): \newline GHZ, QFT mirror, \newline VQE~H$_2$ &
Local server \newline (System~1) &
AMD Ryzen~9 9950X, \newline 16C/32T, \newline 128\,GB DDR5 &
NVIDIA RTX~6000 Ada, \newline 48\,GB VRAM, \newline 18{,}176 CUDA cores \\
\addlinespace
Phase~2 \newline (Exp.~4): \newline VQE UCCSD vs.\ \newline LCNot-UCCSD &
Local server \newline (System~1) and \newline NVIDIA-IHUB \newline infrastructure$^{\dagger}$ &
AMD Ryzen~9 9950X, \newline 16C/32T, \newline 128\,GB DDR5 (System~1) &
NVIDIA DGX~A100 \newline (1$\times$ A100), \newline 80\,GB VRAM, \newline 6{,}912 CUDA cores \\
\addlinespace
Phase~2--3 \newline (Exp.~5--6): \newline H-chain sampling, \newline N$_2$ PES scan &
High-end workstation \newline (System~2) &
AMD Ryzen Threadripper \newline PRO 9975WX, \newline 32C/64T, 256\,GB &
NVIDIA RTX~PRO~6000 \newline Blackwell Workstation \newline Max-Q, 96\,GB VRAM, \newline 24{,}064 CUDA cores \\
\bottomrule
\end{tabular}

\vspace{2pt}
\footnotesize $^{\dagger}$ GPU access for Experiment~4 was provided by the NVIDIA-IHUB ``Challenge to be Quantum Ready'' programme; a single A100 device of the DGX~A100 node was used. The Qiskit/Aer baseline for Experiment~4 was executed on the System~1 host CPU.
\end{table*}

\subsection{Electronic-Structure Preliminaries and Notation}
\label{subsec:prelim}

For the chemistry experiments (Phases~2 and~3), each molecular Hamiltonian is obtained within the Born--Oppenheimer approximation~\cite{Born1927, AKSV2026} from a restricted Hartree--Fock (RHF) reference~\cite{Hartree1928, Fock1930, Roothaan1951} in a chosen Gaussian basis, and mapped to qubits by the Jordan--Wigner (JW) transformation~\cite{Jordan1928, AKSV2026}. An active space is specified by its number of spatial orbitals $N_{\mathrm{orb}}$ and active electrons $N_{\mathrm{ele}}$; under JW, this requires
\begin{equation}
  N_Q = 2\,N_{\mathrm{orb}}
  \label{eq:qubit_count}
\end{equation}
qubits, since each spatial orbital contributes two spin orbitals. We use the notation $(N_{\mathrm{orb}}, N_{\mathrm{ele}})$ throughout and state the qubit count explicitly; for example, a $(16,14)$ active space corresponds to $N_Q = 32$ qubits. For a closed-shell singlet, the particle-number- and $S_z$-conserving (symmetry) subspace has dimension
\begin{equation}
  |\mathbb{S}| = \binom{N_{\mathrm{orb}}}{N_{\mathrm{ele}}/2}^{\!2},
  \label{eq:sym_space}
\end{equation}
which fixes the classical diagonalization target in the sampling-based experiments. Where variational parameters are initialised perturbatively, the second-order M{\o}ller--Plesset (MP2)~\cite{Moller1934} double-excitation amplitudes
\begin{equation}
  t_{ij}^{ab} = \frac{\langle ij \| ab \rangle}{\epsilon_i + \epsilon_j - \epsilon_a - \epsilon_b}
  \label{eq:mp2}
\end{equation}
are used, whose evaluation from stored two-electron integrals scales as $\mathcal{O}(N^4)$ in the system-size measure $N$. Reference energies for accuracy assessment are taken from RHF~\cite{Hartree1928 , Fock1930}, CCSD~\cite{bartlett_2007_CC}, and FCI~\cite{SzaboOstlund1996_fci} computed with PySCF.

\subsection{Phase 1: Foundational Simulator Benchmarks}
\label{subsec:phase1}

The three Phase-1 experiments share a common measurement harness so that framework behaviour can be read on a like-for-like basis. In every case, the CUDA-Q kernel is executed directly on the NVIDIA backend in its constructed form, whereas the Qiskit circuit is passed through the preset pass manager at optimisation level~3 (the highest available) before execution on Aer; this asymmetry is intentional and is accounted for by separating transpilation time from execution time in the timing model (Section~\ref{subsec:protocol}). The CUDA-Q GPU path is evaluated at both single-precision (fp32) and double-precision (fp64), since arithmetic precision is a primary axis of the study. In contrast, the Qiskit statevector path serves as the CPU reference and is computed by default in double precision.

\subsubsection{Experiment 1: GHZ-state preparation}
\label{subsubsec:exp1}
\begin{figure}[htbp]
\centering
\begin{quantikz}[column sep=0.4cm]
\lstick{$q_0$} & \gate{H} & \ctrl{1} & \qw      & \qw      & \meter{} \\
\lstick{$q_1$} & \qw      & \targ{}  & \ctrl{1} & \qw      & \meter{} \\
\lstick{$q_2$} & \qw      & \qw      & \targ{}  & \ctrl{1} & \meter{} \\
\lstick{$q_3$} & \qw      & \qw      & \qw      & \targ{}  & \meter{}
\end{quantikz}
\caption{Linear GHZ-state preparation, shown for $n=4$: a Hadamard on $q_0$ followed by a nearest-neighbour CNOT ladder and terminal measurement. The construction generalises to arbitrary $n$ with backend-agnostic depth $n$.}
\label{fig:ghz}
\end{figure}
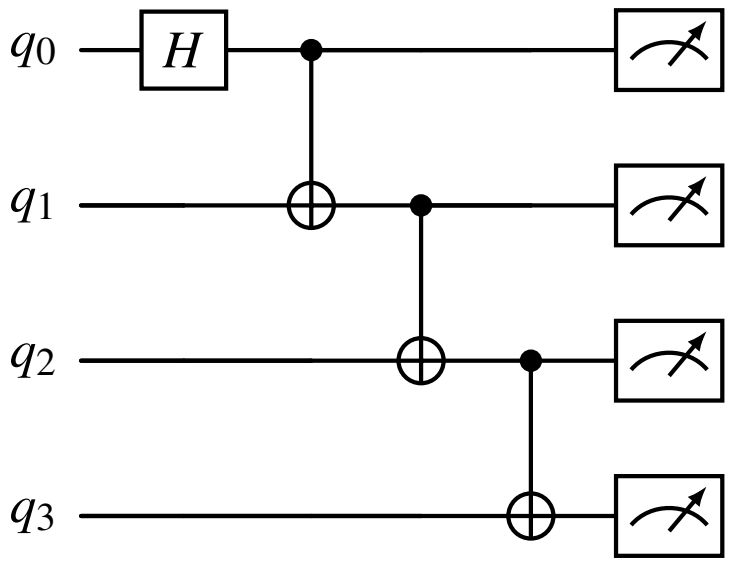

The $n$-qubit Greenberger--Horne--Zeilinger state~\cite{Greenberger2007, Dik1999} is prepared with the canonical linear construction: a Hadamard on the first qubit followed by a ladder of $n-1$ nearest-neighbour CNOTs, giving a backend-agnostic depth of $n$ (Figure~\ref{fig:ghz}). In Qiskit, the circuit uses native \texttt{h} and \texttt{cx} gates with a terminal \texttt{measure\_all}; in CUDA-Q, the equivalent kernel allocates an $n$-qubit register and applies \texttt{h} followed by the CNOT ladder, measured with \texttt{mz}. The measured distribution provides a direct correctness signature as the ideal output populates only the all-zeros and all-ones bitstrings with equal weight. The experiment isolates entangling-gate throughput and width scaling in the absence of parameterized rotations, establishing the baseline against which the deeper Phase-1 circuits are read.

\subsubsection{Experiment 2: QFT--QFT$^{\dagger}$ mirror circuits}
\label{subsubsec:exp2}
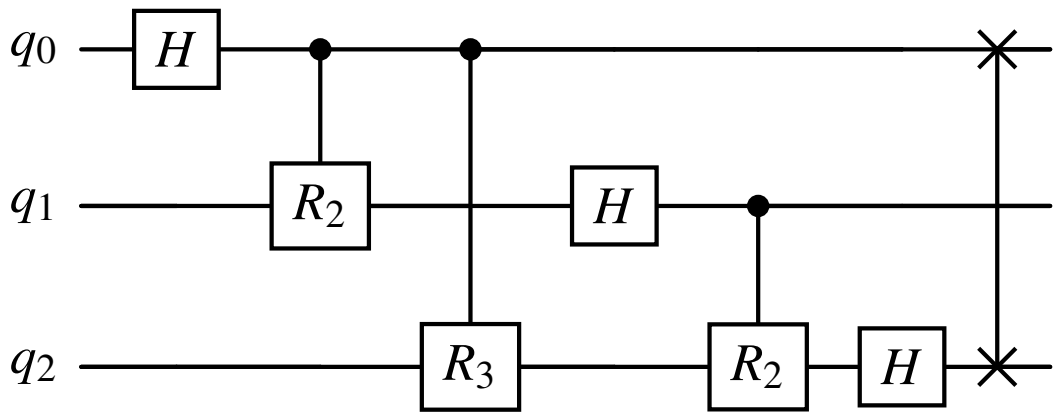
\begin{figure}[htbp]
\centering
\begin{quantikz}[column sep=0.35cm]
\lstick{$q_0$} & \gate{H} & \ctrl{1}   & \ctrl{2}   & \qw      & \qw        & \qw      & \swap{2} & \qw \\
\lstick{$q_1$} & \qw      & \gate{R_2} & \qw        & \gate{H} & \ctrl{1}   & \qw      & \qw      & \qw \\
\lstick{$q_2$} & \qw      & \qw        & \gate{R_3} & \qw      & \gate{R_2} & \gate{H} & \targX{} & \qw
\end{quantikz}
\caption{Gate-level primitive QFT for $n=3$: Hadamards, controlled-phase rotations $R_k$ (a phase of $2\pi/2^{k}$, so $R_k$ at $q_i$--$q_j$ has $k=(j-i)+1$), and a bit-reversal SWAP. The inverse block mirrors this ordering with negated phase angles.}
\label{fig:qft-gatelevel}
\end{figure}

\begin{figure}[htbp]
\centering
\begin{quantikz}[column sep=0.5cm]
\lstick{$q_0$} & \gate[4]{\mathrm{QFT}}\gategroup[4,steps=2,style={dashed, rounded corners, inner sep=6pt}, background, label style={label position=below, yshift=-0.55cm}]{$\times\, r_{\text{reps}}$} & \gate[4]{\mathrm{QFT}^{\dagger}} & \meter{} & \rstick{$\ket{0}$} \\
\lstick{$q_1$} &  &  & \meter{} & \rstick{$\ket{0}$} \\
\lstick{$q_2$} &  &  & \meter{} & \rstick{$\ket{0}$} \\
\lstick{$q_3$} &  &  & \meter{} & \rstick{$\ket{0}$}
\end{quantikz}
\caption{QFT--QFT$^{\dagger}$ mirror circuit on $n$ qubits, shown for $n=4$. The transform and its inverse are applied in sequence and repeated $r_{\text{reps}}$ times. The composition reduces to the identity on all computational-basis states, so the ideal measured bitstring is $\ket{0}^{\otimes n}$.}
\label{fig:qft-mirror-circuit}
\end{figure}
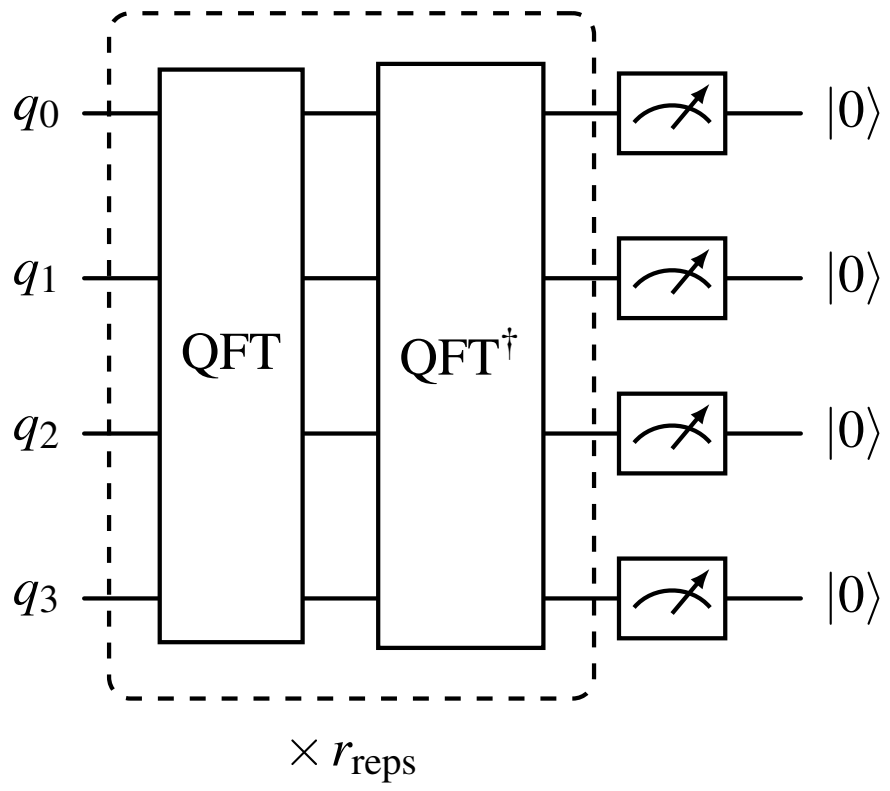

The second experiment stresses resource scaling with a mirror circuit that composes the quantum Fourier transform with its inverse, reducing to the identity on all computational-basis states. The transform is built from primitives rather than a library call: for each qubit $i$, a Hadamard is applied, followed by controlled-phase rotations $R(\pi/2^{\,j-i})$ coupling $i$ to every $j>i$, and a bit-reversal layer of $\lfloor n/2 \rfloor$ SWAPs (Figure~\ref{fig:qft-gatelevel}); the inverse block reverses the gate order with negated rotation angles, so that the full mirror composition reduces to the identity (Figure~\ref{fig:qft-mirror-circuit}). The Qiskit implementation uses \texttt{h}, \texttt{cp}, and \texttt{swap}; the CUDA-Q kernel uses \texttt{h}, \texttt{r1.ctrl}, and \texttt{swap} with the controlled-phase sign inverted for the inverse block. Widths from 1 to 31 qubits are evaluated at shot counts of $10{,}000$ and $100{,}000$ and at circuit repetition counts of 1 and 10, so that width-driven and depth-driven scaling can be separated. Each configuration is run for 10 timed trials preceded by an untimed warm-up. Correctness was verified per trial against the all-zeros bitstring, since $\mathrm{QFT}^{\dagger}(\mathrm{QFT}(\ket{0}^{\otimes n})) = \ket{0}^{\otimes n}$.

Resource accounting differs by necessity between the two pipelines, and this asymmetry is reported explicitly. For Qiskit, gate counts and depth are read from the circuit object both before and after transpilation via \texttt{count\_ops()} and \texttt{depth()}, exposing compiler-induced overhead directly. For CUDA-Q, where no transpiled-circuit object is materialised, gate counts and depth are computed analytically from the closed-form QFT structure ($2n$ Hadamards, $n(n-1)$ controlled-phase rotations, and $2\lfloor n/2 \rfloor$ SWAPs per repetition), and these analytical estimates are cross-checked against the pre-transpiled Qiskit circuit for consistency.

\subsubsection{Experiment 3: VQE on H$_2$ (CUDA-Q versus CUDA-QX)}
\label{subsubsec:exp3}
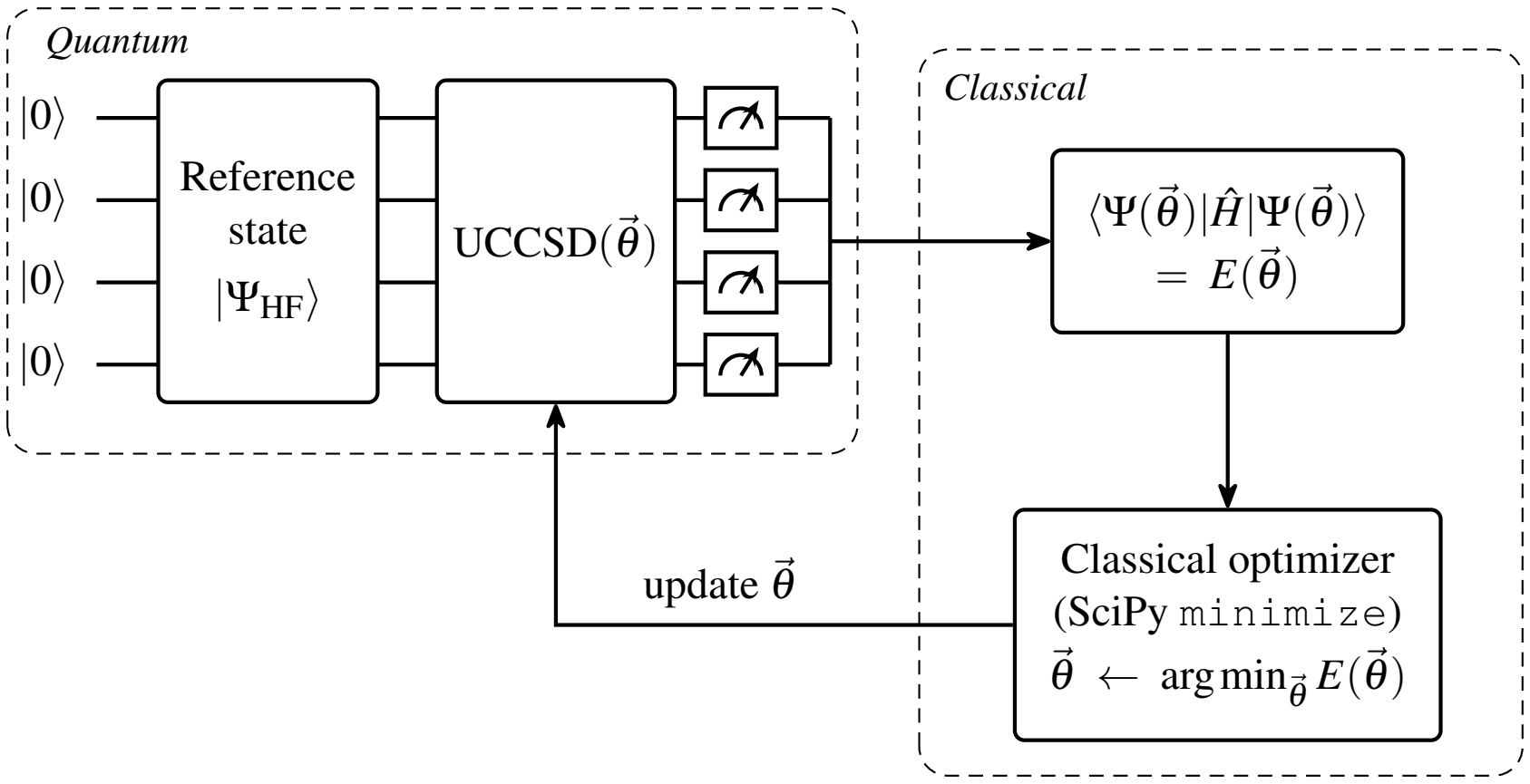
\begin{figure}[htbp]
\centering
\begin{tikzpicture}[
  >={Stealth[round]},
  font=\small,
  blk/.style={draw, thick, rounded corners=2pt, fill=white, align=center},
  est/.style={draw, thick, rounded corners=2pt, fill=white, align=center, inner sep=7pt},
  reglbl/.style={font=\footnotesize\itshape}
]

% ---------------- qubit wires ----------------
\foreach \y in {3.35,2.75,2.15,1.55}{
  \draw[thick] (0.70,\y) -- (5.14,\y);
  \node[left] at (0.62,\y) {$\ket{0}$};
}

% ---------------- reference-state and ansatz blocks ----------------
\node[blk, minimum width=1.6cm, minimum height=2.35cm] (ref) at (1.95,2.45)
  {Reference\\ state\\[3pt] $\ket{\Psi_{\mathrm{HF}}}$};
\node[blk, minimum width=1.6cm, minimum height=2.35cm] (ucc) at (4.05,2.45)
  {$\mathrm{UCCSD}(\vec{\theta})$};

% ---------------- measurement meters ----------------
\foreach \y in {3.35,2.75,2.15,1.55}{
  \begin{scope}[shift={(5.40,\y)}]
    \draw[thick, fill=white] (-0.26,-0.22) rectangle (0.26,0.22);
    \draw[thick] (-0.15,-0.08) arc (180:0:0.15);
    \draw[thick,-{Stealth[length=4pt]}] (0,-0.08) -- (0.13,0.13);
  \end{scope}
}

% ---------------- readout bus ----------------
\foreach \y in {3.35,2.75,2.15,1.55}{ \draw[thick] (5.66,\y) -- (6.05,\y); }
\draw[thick] (6.05,1.55) -- (6.05,3.35);

% ---------------- energy estimator ----------------
\node[est] (E) at (8.95,2.45)
   {$\langle\Psi(\vec{\theta})|\hat{H}|\Psi(\vec{\theta})\rangle$\\[2pt]$=\ E(\vec{\theta})$};

% ---------------- classical optimizer ----------------
\node[est] (O) at (8.95,-0.35)
   {Classical optimizer\\ (SciPy \texttt{minimize})\\[2pt]
    $\vec{\theta}\ \leftarrow\ \arg\min_{\vec{\theta}} E(\vec{\theta})$};

% ---------------- region demarcation ----------------
\draw[densely dashed, rounded corners=6pt] (0.05,0.90) rectangle (6.25,4.15);
\node[reglbl] at (0.85,3.90) {Quantum};

\draw[densely dashed, rounded corners=6pt] (6.70,-1.45) rectangle (11.10,3.85);
\node[reglbl] at (7.40,3.58) {Classical};

% ---------------- signal flow ----------------
\draw[->, thick] (6.05,2.45) -- (E.west);
\draw[->, thick] (E.south) -- (O.north);
\draw[->, thick] (O.west) -- (4.05,-0.35) -- (ucc.south);
\node[above] at (5.25,-0.35) {update $\vec{\theta}$};

\end{tikzpicture}
\caption{Hybrid quantum--classical loop of the VQE used in Experiment~3, shown for the four-qubit H$_2$ Hamiltonian. A reference-state block prepares the Hartree--Fock determinant $\ket{\Psi_{\mathrm{HF}}}$ and the UCCSD$(\vec{\theta})$ ansatz generates the trial state $|\Psi(\vec{\theta})\rangle$; upon measurement the quantum stage completes, and the energy $E(\vec{\theta})=\langle\Psi(\vec{\theta})|\hat{H}|\Psi(\vec{\theta})\rangle$ is evaluated classically from the measured distribution. The classical optimizer then updates $\vec{\theta}$ and the loop repeats until convergence. The dashed regions separate the quantum state-preparation and measurement stages from the classical expectation-value evaluation and parameter update.}
\label{fig:vqe-loop}
\end{figure}

The third experiment introduces a variational chemistry workload while keeping the system minimal, so that the two CUDA-Q programming routes can be compared without the confound of a large Hamiltonian (Figure~\ref{fig:vqe-loop}). Both routes prepare the HF reference with $X$ gates on the occupied qubits and apply a UCCSD ansatz~\cite{Peruzzo2014, Barkoutsos2018, Romero_2019} to the H$_2$ Hamiltonian (JW-mapped from an OpenFermion/PySCF molecule). The first route (``CUDA-Q'') builds the ansatz with the platform's built-in \texttt{cudaq.kernels.uccsd} primitive and evaluates the energy with \texttt{cudaq.observe}, driving the optimisation with SciPy's \texttt{minimize}. The second route (``CUDA-QX'') builds the same ansatz through the \texttt{cudaq-solvers} state-preparation library (\texttt{stateprep.uccsd}) and optimises through the library's own \texttt{solvers.vqe} entry point, again with SciPy as the classical optimiser. Both routes share identical initial parameters, optimiser method, and tolerance, so the comparison measures the overhead and convergence behaviour of the two software abstractions on the same physics.

\subsubsection{The LCNot-UCCSD ansatz and its approximation hierarchy}
\label{subsubsec:lcnot}
The linear-CNOT UCCSD (LCNot-UCCSD) ansatz~\cite{Magoulas2023,Magoulas2023b, KSV2026} replaces the Jordan--Wigner parity chains of standard UCCSD with qubit-excitation-based (QEB) operators, so that each excitation is realised with a CNOT count that scales linearly in its many-body rank rather than exponentially. The trial state factorises the singles and doubles as
\begin{equation}
  \ket{\Psi_{\mathrm{LC}}(\bm{\theta})}
  = \prod_{\substack{i<j\\ a<b}} e^{\theta_{ijab}\,\hat{\tau}^{(2)}_{ijab}}
    \prod_{i,a} e^{\theta_{ia}\,\hat{\tau}^{(1)}_{ia}}\,
    \ket{\Phi_{\mathrm{HF}}},
  \label{eq:lcnot}
\end{equation}
with anti-Hermitian qubit-excitation generators $\hat{\tau}^{(1)}$ and $\hat{\tau}^{(2)}$ acting on occupied/virtual spin-orbital indices. In the implementation used here, a single excitation on the qubit pair $(q_a,q_i)$ is a two-CNOT, one-controlled-$R_y$ sequence, $\mathrm{CX}(q_a,q_i)$--$\mathrm{CR}_y(\theta)$--$\mathrm{CX}(q_a,q_i)$, and a double excitation on the quartet $(q_a,q_b,q_i,q_j)$ uses a three-CNOT ladder followed by open-control $X$ gates, a triply controlled $R_y$ (CCCRy) applied across the four qubits, and the inverse sequence to restore the register (Figure~\ref{fig:lcnot-double}). In the CUDA-Q kernel the CCCRy is expressed natively (\texttt{ry.ctrl} with a multi-qubit control list) and is simulated as a genuine multi-qubit unitary, whereas the Qiskit construction realises it as a \texttt{RYGate($\theta$).control(3)}; the Qiskit LCNot-UCCSD path maps the fermionic operator with an \emph{interleaved} JW ordering (\texttt{InterleavedQubitMapper}) to match the spin-orbital layout of the custom circuit.

The dominant source of two-qubit gates in these circuits is the multi-controlled $R_y$ rotation. Following the parent construction~\cite{Magoulas2023}, controls may be removed from this rotation to reduce the CNOT count further, at the cost of breaking particle-number ($N$) and $S_z$ symmetry while preserving spatial symmetry. The implementation exposes this as a discrete hierarchy of double-excitation approximation levels (AL0--AL3), together with an optional single-excitation approximation, as shown in Figure~\ref{fig:lcnot-hierarchy}. AL0 is the exact QEB double excitation; AL1 and AL2 progressively drop controls; AL3 replaces the controlled rotation with a bare single-qubit $R_y$, the most aggressive reduction. The per-excitation gate model of the resource estimator (used to populate the resource tables in Section~\ref{sec:results}) is fixed by this level: every level carries a common six-CNOT ladder, and the levels differ in the number of open-control $X$ gates and in the arity of the retained $R_y$ rotation.

\begin{figure}[htbp]
\centering
\begin{quantikz}[column sep=0.32cm, row sep=0.30cm]
\lstick{$q_a$} & \ctrl{1} & \ctrl{2} & \qw       & \gate{R_y(\theta)} & \qw       & \ctrl{2} & \ctrl{1} & \qw \\
\lstick{$q_b$} & \targ{}  & \qw      & \gate{X}  & \control{}         & \gate{X}  & \qw      & \targ{}  & \qw \\
\lstick{$q_i$} & \ctrl{1} & \targ{}  & \qw       & \control{}         & \qw       & \targ{}  & \ctrl{1} & \qw \\
\lstick{$q_j$} & \targ{}  & \qw      & \gate{X}  & \ctrl{-3}          & \gate{X}  & \qw      & \targ{}  & \qw
\end{quantikz}
\caption{Exact (AL0) double-excitation circuit of the LCNot-UCCSD ansatz on the spin-orbital quartet $(q_a,q_b,q_i,q_j)$. A three-CNOT entangling ladder is followed by open-control $X$ gates on $q_b$ and $q_j$, a triply-controlled rotation $R_y(\theta)$ with target $q_a$ and controls $\{q_b,q_i,q_j\}$, and the mirrored inverse ladder that restores the register. The construction uses six CNOTs, four $X$ gates, and one CCCRy.}
\label{fig:lcnot-double}
\end{figure}
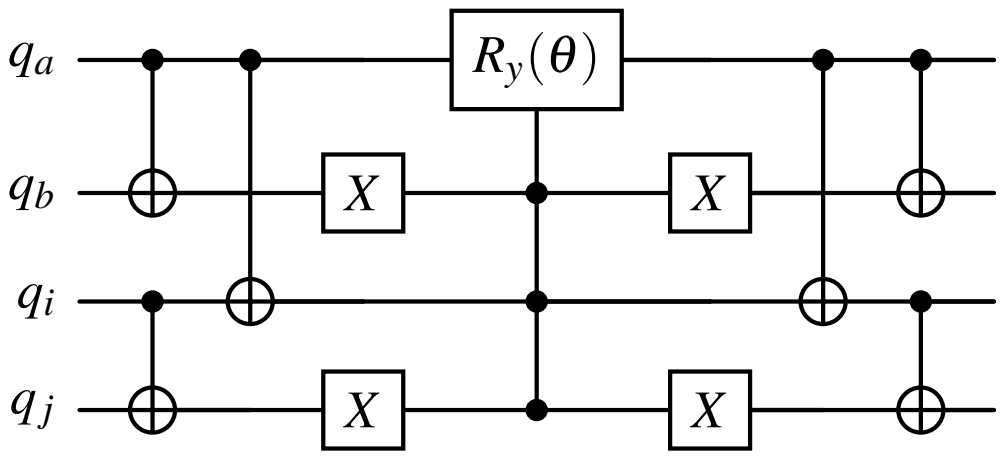

\begin{figure}[htbp]
\centering
\begin{subfigure}[b]{0.46\textwidth}\centering
\makebox[\linewidth]{%
\begin{quantikz}[row sep={0.62cm,between origins}, column sep=0.45cm]
\lstick{$q_a$} & \qw & \qw      & \gate{R_y(\theta)} & \qw      & \qw & \qw  \\
\lstick{$q_b$} & \qw & \gate{X} & \control{}         & \gate{X} & \qw & \qw  \\
\lstick{$q_i$} & \qw & \qw      & \control{}         & \qw      & \qw & \qw  \\
\lstick{$q_j$} & \qw & \gate{X} & \ctrl{-3}          & \gate{X} & \qw & \qw 
\end{quantikz}}
\caption{AL0: CCCRy, three controls, four $X$.}
\end{subfigure}\hfill
\begin{subfigure}[b]{0.46\textwidth}\centering
\makebox[\linewidth]{%
\begin{quantikz}[row sep={0.62cm,between origins}, column sep=0.45cm]
\lstick{$q_a$} & \qw & \qw      & \gate{R_y(\theta)} & \qw      & \qw & \qw  \\
\lstick{$q_b$} & \qw & \qw      & \qw                & \qw      & \qw & \qw  \\
\lstick{$q_i$} & \qw & \qw      & \control{}         & \qw      & \qw & \qw  \\
\lstick{$q_j$} & \qw & \gate{X} & \ctrl{-3}          & \gate{X} & \qw & \qw 
\end{quantikz}}
\caption{AL1: CCRy, two controls, two $X$.}
\end{subfigure}

\vspace{1.0em}

\begin{subfigure}[b]{0.46\textwidth}\centering
\makebox[\linewidth]{%
\begin{quantikz}[row sep={0.62cm,between origins}, column sep=0.45cm]
\lstick{$q_a$} & \qw & \qw      & \gate{R_y(\theta)} & \qw      & \qw & \qw \\
\lstick{$q_b$} & \qw & \qw      & \qw                & \qw      & \qw & \qw \\
\lstick{$q_i$} & \qw & \qw      & \qw                & \qw      & \qw & \qw \\
\lstick{$q_j$} & \qw & \gate{X} & \ctrl{-3}          & \gate{X} & \qw & \qw
\end{quantikz}}
\caption{AL2: CRy, one control, two $X$.}
\end{subfigure}\hfill
\begin{subfigure}[b]{0.46\textwidth}\centering
\makebox[\linewidth]{%
\begin{quantikz}[row sep={0.62cm,between origins}, column sep=0.45cm]
\lstick{$q_a$} & \qw & \qw & \gate{R_y(\theta)} & \qw & \qw & \qw \\
\lstick{$q_b$} & \qw & \qw & \qw                & \qw & \qw & \qw \\
\lstick{$q_i$} & \qw & \qw & \qw                & \qw & \qw & \qw  \\
\lstick{$q_j$} & \qw & \qw & \qw                & \qw & \qw & \qw 
\end{quantikz}}
\caption{AL3: bare $R_y$, no controls, no $X$.}
\end{subfigure}
\caption{Interchangeable central rotation block of the LCNot-UCCSD double excitation across approximation levels AL0--AL3. All four levels share the six-CNOT ladder of Figure~\ref{fig:lcnot-double}; successively removing controls from the rotation lowers the $X$ and effective two-qubit cost while breaking particle-number and $S_z$ symmetry.}
\label{fig:lcnot-hierarchy}
\end{figure}
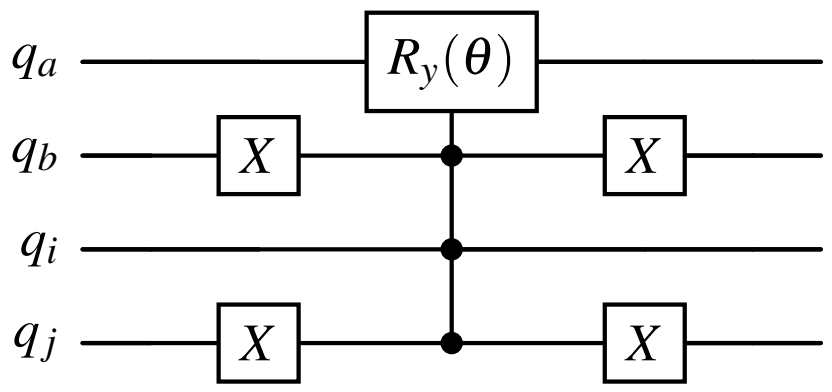
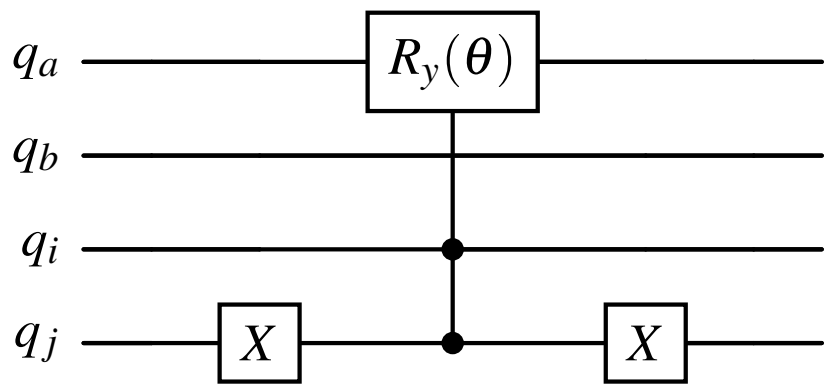
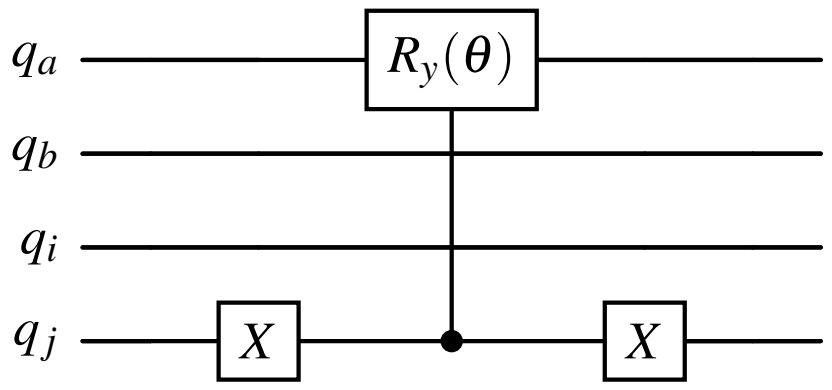
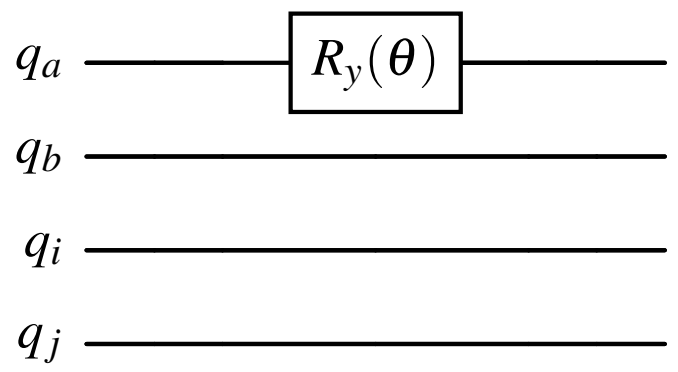

\subsection{Phase 2: Quantum Chemistry Algorithm Benchmarks}
\label{subsec:phase2}
\subsubsection{Experiment 4: UCCSD versus LCNot-UCCSD VQE across molecules}
\label{subsubsec:exp4}
\begin{figure*}[htbp]
\centering
\newcommand{\moLev}[2]{\draw[thick] (#1-0.42,#2) -- (#1+0.42,#2);}
\newcommand{\moArr}[2]{%
  \draw[-{Stealth[length=3pt]}] (#1-0.12,#2-0.14) -- (#1-0.12,#2+0.14);
  \draw[-{Stealth[length=3pt]}] (#1+0.12,#2+0.14) -- (#1+0.12,#2-0.14);}
\newcommand{\moFroz}[2]{\fill[black!12, rounded corners=1pt] (#1-0.52,#2-0.17) rectangle (#1+0.52,#2+0.17);}
\newcommand{\moBrace}[4]{% x ybot ytop label
  \draw[decorate, decoration={brace, amplitude=4pt, mirror}, thin]
    (#1+0.5,#2-0.17) -- (#1+0.5,#3+0.17) node[midway, right=5pt]{\scriptsize #4};}
\newcommand{\moName}[3]{% x name detail
  \node at (#1,-0.55) {#2};
  \node at (#1,-0.94) {\scriptsize #3};}
\begin{tikzpicture}[font=\footnotesize]

% ---------------- H2 (x=0) : 0 frozen, 1 occ, 1 virt ----------------
\moLev{0}{0}\moArr{0}{0}
\moLev{0}{0.5}
\moBrace{0}{0}{0.5}{4 qubits}
\moName{0}{H$_2$}{$N_{\mathrm{orb}}{=}2,\ N_{\mathrm{ele}}{=}2$}

% ---------------- LiH (x=2.8) : 1 frozen, 1 occ, 4 virt ----------------
\moFroz{2.8}{0}\moLev{2.8}{0}\moArr{2.8}{0}
\moLev{2.8}{0.5}\moArr{2.8}{0.5}
\moLev{2.8}{1.0}\moLev{2.8}{1.5}\moLev{2.8}{2.0}\moLev{2.8}{2.5}
\moBrace{2.8}{0.5}{2.5}{10 qubits}
\moName{2.8}{LiH}{$N_{\mathrm{orb}}{=}5,\ N_{\mathrm{ele}}{=}2$}

% ---------------- HF (x=5.6) : 1 frozen, 4 occ, 1 virt ----------------
\moFroz{5.6}{0}\moLev{5.6}{0}\moArr{5.6}{0}
\moLev{5.6}{0.5}\moArr{5.6}{0.5}
\moLev{5.6}{1.0}\moArr{5.6}{1.0}
\moLev{5.6}{1.5}\moArr{5.6}{1.5}
\moLev{5.6}{2.0}\moArr{5.6}{2.0}
\moLev{5.6}{2.5}
\moBrace{5.6}{0.5}{2.5}{10 qubits}
\moName{5.6}{HF}{$N_{\mathrm{orb}}{=}5,\ N_{\mathrm{ele}}{=}8$}

% ---------------- BeH2 (x=8.4) : 1 frozen, 2 occ, 4 virt ----------------
\moFroz{8.4}{0}\moLev{8.4}{0}\moArr{8.4}{0}
\moLev{8.4}{0.5}\moArr{8.4}{0.5}
\moLev{8.4}{1.0}\moArr{8.4}{1.0}
\moLev{8.4}{1.5}\moLev{8.4}{2.0}\moLev{8.4}{2.5}\moLev{8.4}{3.0}
\moBrace{8.4}{0.5}{3.0}{12 qubits}
\moName{8.4}{BeH$_2$}{$N_{\mathrm{orb}}{=}6,\ N_{\mathrm{ele}}{=}4$}

% ---------------- H2O (x=11.2) : 1 frozen, 4 occ, 2 virt ----------------
\moFroz{11.2}{0}\moLev{11.2}{0}\moArr{11.2}{0}
\moLev{11.2}{0.5}\moArr{11.2}{0.5}
\moLev{11.2}{1.0}\moArr{11.2}{1.0}
\moLev{11.2}{1.5}\moArr{11.2}{1.5}
\moLev{11.2}{2.0}\moArr{11.2}{2.0}
\moLev{11.2}{2.5}\moLev{11.2}{3.0}
\moBrace{11.2}{0.5}{3.0}{12 qubits}
\moName{11.2}{H$_2$O}{$N_{\mathrm{orb}}{=}6,\ N_{\mathrm{ele}}{=}8$}

% ---------------- legend (single line) ----------------
\moFroz{0}{-1.7}\moLev{0}{-1.7}\moArr{0}{-1.7}
\node[right] at (0.6,-1.7) {Frozen core (excluded)};
\moLev{4.5}{-1.7}\moArr{4.5}{-1.7}
\node[right] at (5.1,-1.7) {Active, occupied in HF};
\moLev{9.1}{-1.7}
\node[right] at (9.7,-1.7) {Active, virtual in HF};
\end{tikzpicture}
\caption{Schematic active-space partition for the five Experiment~4 molecules in the STO-3G basis (energy ordering illustrative, not to scale). Each horizontal line is one spatial orbital, mapping to two spin orbitals and hence two qubits. Shaded levels are the frozen core (doubly occupied and removed by the frozen-core transformation for all molecules except H$_2$); unshaded levels form the active space, with occupancy shown for the closed-shell Hartree--Fock reference. The active space extends to the top of the STO-3G orbital set in every case, so no orbitals are excluded as frozen virtuals; the right-hand brace gives the resulting qubit register width.}
\label{fig:active-spaces}
\end{figure*}
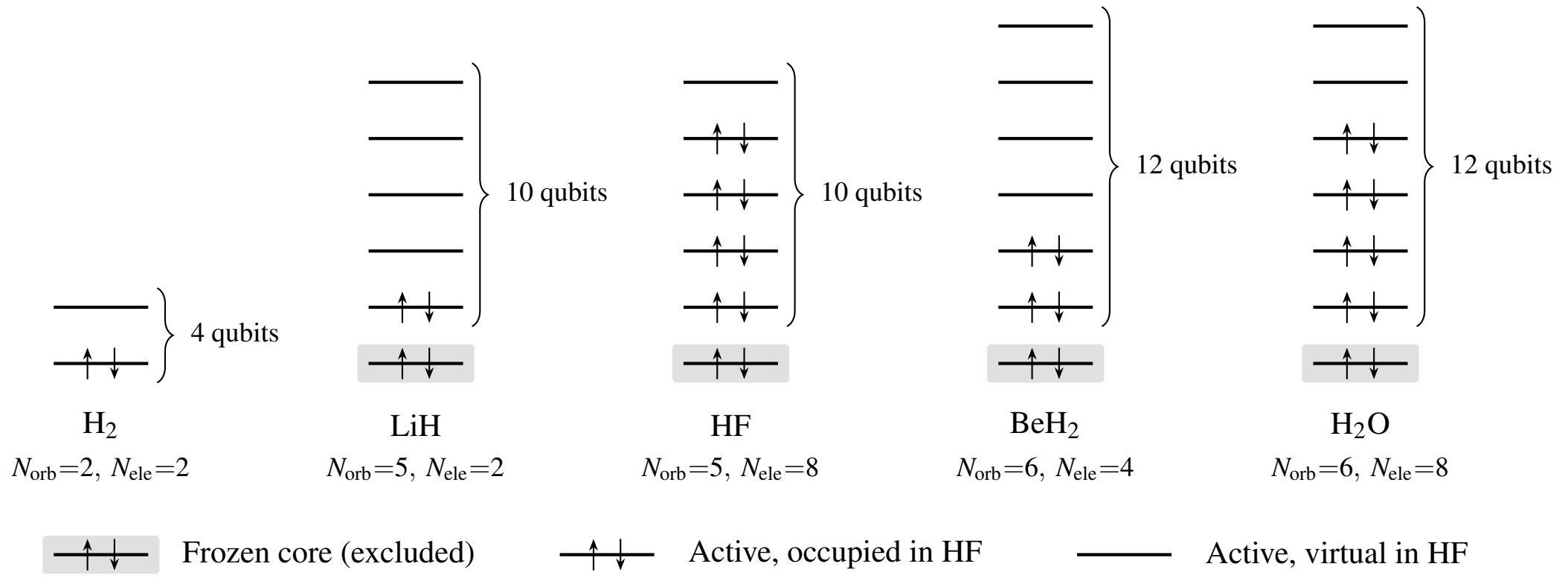

Experiment~4 compares the standard UCCSD ansatz against LCNot-UCCSD on five molecules, H$_2$, LiH, HF, BeH$_2$, and H$_2$O, in the STO-3G basis, under both frameworks, giving four scripted pipelines (Qiskit-UCCSD, Qiskit-LCNot-UCCSD, CUDA-Q-UCCSD, CUDA-Q-LCNot-UCCSD). Active spaces follow a fixed core-freezing scheme: H$_2$ $(2,2)$/4~qubits; LiH $(5,2)$/10~qubits; HF $(5,8)$/10~qubits; BeH$_2$ $(6,4)$/12~qubits; H$_2$O $(6,8)$/12~qubits, with a single frozen core orbital for all but H$_2$ (Figure~\ref{fig:active-spaces}). The classical optimiser is SLSQP with tolerance $10^{-6}$ in every pipeline. The Qiskit path constructs the Hamiltonian through the PySCF driver with a frozen-core transformer, maps it with the Jordan--Wigner mapper (interleaved for LCNot-UCCSD), evaluates energies through the Aer \texttt{EstimatorV2} primitive after level-3 transpilation, and adds the nuclear-repulsion and frozen-core constants to the electronic energy. The CUDA-Q path builds the Hamiltonian through OpenFermion, evaluates energies with \texttt{cudaq.observe} on the \texttt{nvidia} target at fp64, and prepends an untimed GPU warm-up to remove first-call caching bias. Accuracy is assessed against CCSD and FCI computed with PySCF, and resource usage is captured for both ans\"{a}tze by dedicated estimators: an analytic gate-cost model for LCNot-UCCSD (Figure~\ref{fig:lcnot-hierarchy}) and a category-resolved \texttt{count\_ops()} tally for the transpiled Qiskit circuits. For this experiment, the LCNot-UCCSD circuits use the exact double-excitation level (AL0) and exact singles, so that the comparison against standard UCCSD is a like-for-like accuracy contrast and the CNOT-count reduction is measured without introducing symmetry breaking.

\subsubsection{Experiment 5: LCNot-UCCSD quantum sampling on hydrogen chains}
\label{subsubsec:exp5}
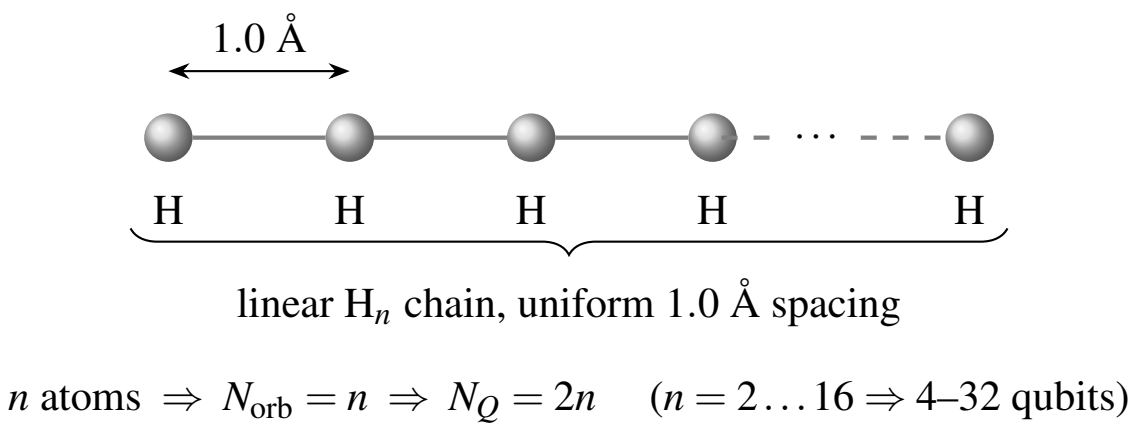
\begin{figure}[htbp]
\centering
\begin{tikzpicture}[font=\footnotesize, >={Stealth[length=4pt]}]
  \def\dx{1.2}
  \foreach \i in {0,1,2,3}{ \shade[ball color=gray!35] (\i*\dx,0) circle (0.16); }
  \node at (4.3,0) {$\cdots$};
  \shade[ball color=gray!35] (5.3,0) circle (0.16);
  % bonds
  \draw[gray, thick] (0.16,0)--(1.04,0);
  \draw[gray, thick] (1.36,0)--(2.24,0);
  \draw[gray, thick] (2.56,0)--(3.44,0);
  \draw[gray, thick, dashed] (3.66,0)--(4.02,0);
  \draw[gray, thick, dashed] (4.58,0)--(5.14,0);
  % atom labels
  \foreach \i in {0,1,2,3}{ \node[below=3pt] at (\i*\dx,-0.16) {H}; }
  \node[below=3pt] at (5.3,-0.16) {H};
  % spacing dimension
  \draw[<->] (0,0.44) -- (1.2,0.44) node[midway, above]{$1.0$~\AA};
  % chain brace
  \draw[decorate, decoration={brace, amplitude=5pt, mirror}, thin]
     (-0.25,-0.60) -- (5.55,-0.60)
     node[midway, below=6pt]{linear H$_n$ chain, uniform $1.0$~\AA{} spacing};
  % scaling map
  \node at (2.65,-1.75)
    {$n$ atoms $\;\Rightarrow\; N_{\mathrm{orb}}=n \;\Rightarrow\; N_Q = 2n$ \quad
     ($n = 2\ldots16 \Rightarrow 4$--$32$ qubits)};
\end{tikzpicture}
\caption{Linear hydrogen-chain geometry for Experiment~5. Chains H$_n$ with $n = 2,\dots,16$ are built with uniform $1.0$~\AA{} nearest-neighbour spacing in STO-3G; the active space comprises $N_{\mathrm{orb}} = n$ spatial orbitals, yielding a register of $N_Q = 2n$ qubits and a sweep from 4 to 32 qubits across the series.}
\label{fig:hchain}
\end{figure}

Experiment~5 shifts from expectation-value estimation to computational-basis \emph{sampling}, the primitive on which the final N$_2$ scan depends, and pushes width to the 32-qubit ceiling. Linear hydrogen chains H$_n$ with uniform spacing of 1.0~\AA{} are constructed for $n = 2,\dots,16$ in STO-3G, giving active spaces of $N_{\mathrm{orb}} = n$ spatial orbitals and hence $N_Q = 2n$ qubits, i.e.\ a sweep from 4 to 32 qubits (Figure~\ref{fig:hchain}). The LCNot-UCCSD circuit is prepared with random initial parameters and sampled for $10{,}000$ shots per trial; the CUDA-Q path samples with \texttt{cudaq.sample} on the \texttt{nvidia} target at fp64 (with an untimed warm-up), while the Qiskit path transpiles at level~3 and samples through the runtime \texttt{SamplerV2} on Aer. Reported quantities include the number of distinct measured bitstrings (the sampled-subspace dimension), sampling time, and memory footprint, alongside the estimated gate counts. 

\begin{figure*}[htbp]
\centering
\begin{tikzpicture}[
  font=\footnotesize, >={Stealth[length=4pt]},
  stage/.style={draw, thick, rounded corners=2pt, align=center, fill=white, inner sep=5pt, minimum height=0.95cm, minimum width=2.5cm},
  lane/.style={font=\small\bfseries, align=right}
]
% Qiskit lane
\node[lane] at (-0.6,2.0) {Qiskit--Aer\\(CPU)};
\node[stage] (q1) at (2.3,2.0) {LCNot-UCCSD\\ AL3 (reduced doubles)};
\node[stage] (q2) at (5.6,2.0) {level-3\\ transpilation};
\node[stage] (q3) at (8.9,2.0) {\texttt{SamplerV2} (Aer)\\ $10^4$ shots};
\draw[->] (q1)--(q2); \draw[->] (q2)--(q3);
% CUDA-Q lane
\node[lane] at (-0.6,0.4) {CUDA-Q\\(GPU)};
\node[stage] (c1) at (2.3,0.4) {LCNot-UCCSD\\ AL0 (exact doubles)};
\node[stage] (c2) at (5.6,0.4) {native kernel\\ (no transpilation)};
\node[stage] (c3) at (8.9,0.4) {\texttt{cudaq.sample}\\ nvidia/fp64, $10^4$ shots};
\draw[->] (c1)--(c2); \draw[->] (c2)--(c3);
\end{tikzpicture}
\caption{Framework-optimal operating points for the Experiment~5 sampling comparison. Each pipeline is configured to minimise its own end-to-end quantum resource cost rather than to a bit-identical circuit: the Qiskit--Aer path uses the fully reduced double-excitation level (AL3) with level-3 transpilation, and the CUDA-Q path uses the exact level (AL0) executed directly as a native multi-qubit kernel without a transpilation stage.}
\label{fig:sampling-config}
\end{figure*}
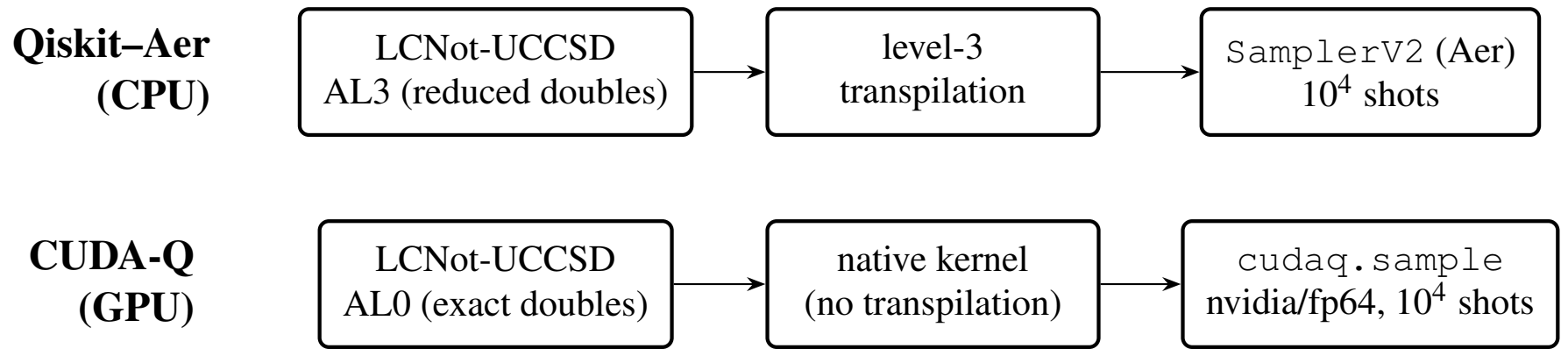

The two sampling pipelines are deliberately configured to their respective best-case operating points for this task rather than to a bit-identical circuit, and the comparison is framed accordingly. The Qiskit--Aer pipeline uses the fully reduced double-excitation level (AL3), which produces the shallowest, lowest-gate-count LCNot-UCCSD circuit, combined with level-3 transpilation, i.e.\ the most favourable compilation the Aer stack offers. The CUDA-Q pipeline uses the exact double-excitation level (AL0) and executes the kernel directly, with native multi-qubit-gate simulation and no transpilation stage. Each framework is therefore timed in the configuration that minimises its own end-to-end sampling cost, and the measurement is a framework-optimal throughput comparison for the sampling primitive rather than a like-for-like fixed-circuit benchmark.The differing approximation levels and the presence or absence of a transpilation stage are an intentional part of this design and are reported alongside the resulting gate counts and depths so that the two operating points are fully specified (Figure~\ref{fig:sampling-config}).

\subsection{Phase 3: Application to the N\texorpdfstring{$_2$}{2} Potential Energy Surface}
\label{subsec:phase3}

\subsubsection{QSCI/SQD and the QSCI-RBM workflow}
\label{subsubsec:qsci}
This final experiment rests on quantum sampling-based classical subspace diagonalization, in which the quantum device (here, the simulator) is used only to identify important electronic configurations, and the Hamiltonian is diagonalised classically within the subspace they span. In Quantum-Selected Configuration Interaction (QSCI)~\cite{Kanno2023}, an ansatz state is measured in the computational basis; the bitstrings that conserve the correct particle number and $S_z$ are retained by post-selection, and the Hamiltonian is projected onto the retained determinants $\mathbb{S}_{\mathrm{sub}}$ and diagonalised with a Davidson eigensolver~\cite{Davidson1975}. Because all matrix elements are evaluated classically via Slater--Condon rules, the energy obeys the variational bound~\cite{MacDonald1933}
\begin{equation}
  E_{\mathrm{QSCI}} = \min\operatorname{eig}\!\big(\hat{H}\big|_{\mathbb{S}_{\mathrm{sub}}}\big) \geq E_{\mathrm{exact}}.
  \label{eq:varbound}
\end{equation}
Sample-based Quantum Diagonalization (SQD)~\cite{Robledo2025} extends QSCI with a configuration-recovery step that probabilistically repairs symmetry-violating bitstrings toward a reference occupancy vector, enlarging $\mathbb{S}_{\mathrm{sub}}$ and improving robustness under noise at the cost of a larger classical diagonalization.

A consequence of the noise-free statevector setting adopted here is worth stating precisely, because it determines what the two baselines actually compute. Under exact statevector simulation the LCNot-UCCSD circuit emits no symmetry-violating bitstrings: every sampled configuration already lies in the particle-number- and $S_z$-conserving sector. QSCI post-selection therefore discards nothing, and the stochastic configuration recovery of SQD has no erroneous samples to repair. The two post-processors are consequently identical in this regime, performing direct diagonalization of the quantum sampled subspace; we refer to this common baseline as SQD throughout. The methodological distinction in this experiment is thus carried entirely by the generative subspace expansion, not by the stochastic configuration recovery of standard SQD.

That subspace expansion paired with traditional QSCI is the QSCI-RBM paradigm~\cite{Patra2026mlcsg}, which augments the post-selected subspace with a compact, learned pool of configurations rather than increasing the configurations stochastically. Symmetry-valid bitstrings seed a monotonically growing determinant memory; a Bernoulli--Bernoulli restricted Boltzmann machine~\cite{Freund1991, Nomura2017, Melko2019RBMQuantumPhysics, Herzog2023GenerativeMLSchrodinger} is trained on that memory; and Gibbs-sampled candidates are symmetry-filtered and merged before re-diagonalization. Because the subspace only grows, the variational bound of Eq.~\eqref{eq:varbound} is preserved at every iteration, and the learned generator concentrates sampling on the high-weight support of the wavefunction. The distinction from the SQD baseline is therefore substantive even in the absence of noise: SQD is bounded by the configurations the finite-shot sampling happens to reveal, whereas QSCI-RBM can recover high-weight determinants that the raw sampling misses.

\subsubsection{Experiment 6: N\texorpdfstring{$_2$}{2} potential-energy-surface scan}
\label{subsubsec:exp6}
This final experiment applies the QSCI(LCNot-UCCSD)--RBM workflow to a potential-energy-surface scan of the nitrogen molecule in the cc-pVDZ basis, using a single active space of 16 spatial orbitals and 10 active electrons, $(N_{\mathrm{orb}},N_{\mathrm{ele}}) = (16,10)$, which by Eq.~\eqref{eq:qubit_count} corresponds to $N_Q = 32$ qubits (the two nitrogen $1s$ cores are frozen, leaving 10 active electrons). This directly extends our group's prior CPU-based statevector work~\cite{KSV2026FX700}, which scanned the same molecule in cc-pVDZ at a smaller $(8,8)$ active space, i.e.\ 16 qubits. The active space is thereby doubled from 8 to 16 spatial orbitals, and correspondingly from 16 to 32 qubits. All the while the algorithmic workflow is held fixed, so that the scan isolates the effect of the enlarged correlation space and the transition to GPU-resident sampling.

\begin{figure}[htbp]
\centering
\newcommand{\nnLev}[2]{\draw[thick] (#1-0.55,#2) -- (#1+0.55,#2);}
\newcommand{\nnArr}[2]{%
  \draw[-{Stealth[length=3pt]}] (#1-0.12,#2-0.13) -- (#1-0.12,#2+0.13);
  \draw[-{Stealth[length=3pt]}] (#1+0.12,#2+0.13) -- (#1+0.12,#2-0.13);}
\begin{tikzpicture}[
  font=\footnotesize,
  lbl/.style={align=right, anchor=east, font=\footnotesize},
  rlbl/.style={align=left, anchor=west, font=\footnotesize}
]
% ---- frozen shaded regions (drawn behind) ----
\fill[black!12, rounded corners=1pt] (-0.85,0.02) rectangle (0.85,0.63);
\fill[black!12, rounded corners=1pt] (-0.85,4.03) rectangle (0.85,5.26);

% ---- frozen core: 2 orb, doubly occupied ----
\nnLev{0}{0.17}\nnArr{0}{0.17}
\nnLev{0}{0.48}\nnArr{0}{0.48}

% ---- active occupied: 5 orb, doubly occupied ----
\foreach \y in {0.92,1.25,1.58,1.91,2.24}{ \nnLev{0}{\y}\nnArr{0}{\y} }

% ---- active virtual: 11 orb (representative, symmetric ellipsis) ----
\foreach \y in {2.60,2.93,3.26}{ \nnLev{0}{\y} }
\node at (0,3.68) {$\vdots$};
\nnLev{0}{3.86}

% ---- frozen virtual: 10 orb (representative, symmetric ellipsis) ----
\nnLev{0}{4.18}
\nnLev{0}{4.51}
\node at (0,4.93) {$\vdots$};
\nnLev{0}{5.11}

% ---- left tier labels ----
\node[lbl] at (-0.98,0.32) {frozen core\\ \scriptsize 2 orb $\cdot$ 4e$^-$};
\node[lbl] at (-0.98,1.58) {active occupied\\ \scriptsize 5 orb $\cdot$ 10e$^-$};
\node[lbl] at (-0.98,3.20) {active virtual\\ \scriptsize 11 orb $\cdot$ 0e$^-$};
\node[lbl] at (-0.98,4.64) {frozen virtual\\ \scriptsize 10 orb $\cdot$ 0e$^-$};

% ---- active-space brace (right, opening toward levels) ----
\draw[decorate, decoration={brace, amplitude=5pt, mirror}, thin]
  (0.95,0.77) -- (0.95,3.99)
  node[midway, rlbl, xshift=7pt]
  {Active space\\ \scriptsize $(N_{\mathrm{orb}},N_{\mathrm{ele}})=(16,10)$\\ \scriptsize $\Rightarrow\ N_Q = 32$ qubits};
\end{tikzpicture}
\caption{Active-space partition of N$_2$ in the cc-pVDZ basis for Experiment~6 (energy ordering schematic, not to scale). Of the 28 spatial orbitals (56 spin orbitals), the two nitrogen $1s$ cores are frozen and doubly occupied, ten valence electrons occupy the five lowest active orbitals in the RHF reference, and eleven active virtual orbitals complete the $(N_{\mathrm{orb}},N_{\mathrm{ele}})=(16,10)$ active space, giving a 32-qubit register. The ten highest cc-pVDZ virtuals are frozen out (excluded from the simulated space). Shaded tiers denote frozen orbitals; the virtual manifolds are drawn with representative levels and orbital counts.}
\label{fig:n2-active-space}
\end{figure}
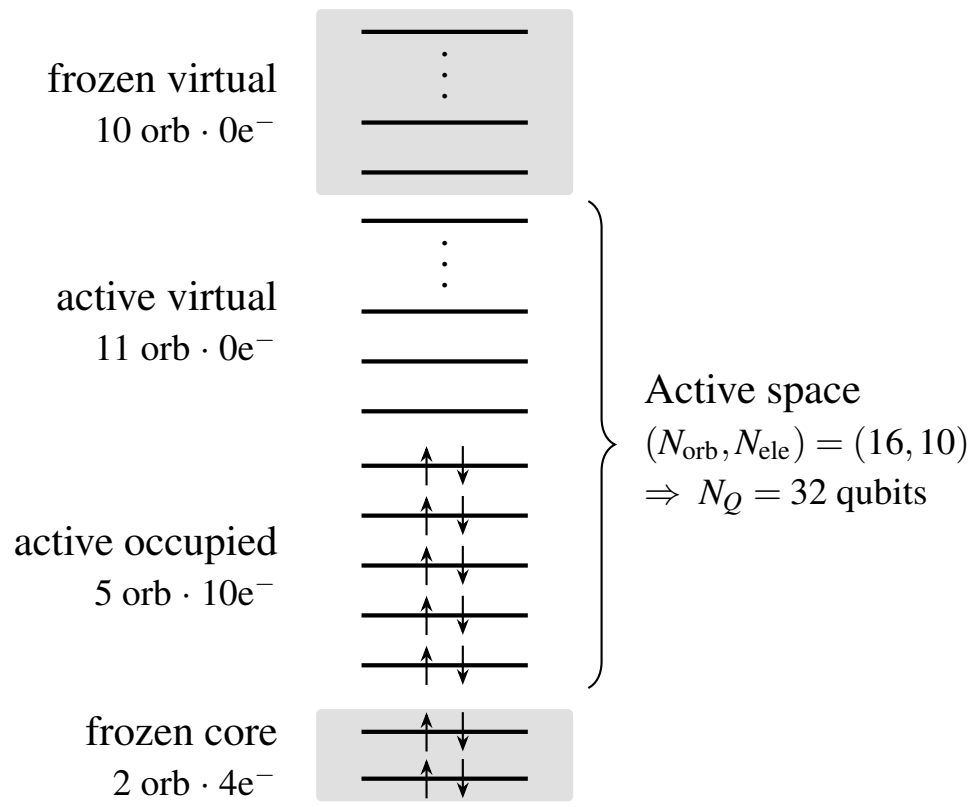

For each internuclear separation on the PES grid, an RHF reference and MP2 amplitudes are computed classically with PySCF; the LCNot-UCCSD circuit, initialised from those amplitudes, is sampled on the RTX~PRO~6000 Blackwell accelerator via the CUDA-Q \texttt{nvidia} target at fp64; and the post-processor builds and diagonalises the subspace. Two post-processors are compared along the scan: the SQD baseline (which, as noted in Section~\ref{subsubsec:qsci}, reduces to direct quantum sampled subspace diagonalization under exact simulation) and QSCI-RBM. Accuracy is assessed against a CASCI reference computed in the same active space, with chemical accuracy threshold (1 kcal/mol) as the deviation threshold.

\subsection{Benchmarking Protocol, Metrics, and Resource Estimation}
\label{subsec:protocol}

A single protocol governs every timed experiment so that measurements are comparable across frameworks, phases, and machines. Each configuration is executed for repeated trials (10 timed trials for the GHZ (Section~\ref{subsubsec:exp1}), QFT mirror circuit benchmarks (Section~\ref{subsubsec:exp2}), 1000 trails for the molecular H$_2$ VQE (Section~\ref{subsubsec:exp3}), 10 timed trails for the ansatz comparison of VQE (Section~\ref{subsubsec:exp4}), 10 sampling trials for the hydrogen-chain (Section~\ref{subsubsec:exp5}) and one run for the production N$_2$ PES workload (Section~\ref{subsubsec:exp6})), and every GPU workload is preceded by an untimed warm-up execution of the same circuit to remove first-call kernel-compilation and caching bias. Reproducibility is enforced by fixing the NumPy random seed. Wall-clock time is decomposed by stage rather than reported as a single number: preprocessing (Hamiltonian construction and, where relevant, parameter initialisation), transpilation (Qiskit only), the timed quantum operation (expectation evaluation for VQE or shot sampling for the sampling experiments), and classical post-processing (reference-energy computation or subspace diagonalization). The process memory footprint is recorded per trial, and GPU memory and utilisation are tracked through \texttt{pynvml}. Reported runtimes are the arithmetic mean over the timed trials, with minimum and maximum shown on a logarithmic scale where dispersion is material.

Resource estimation is framework-appropriate where, for Qiskit circuits, a category-resolved estimator reads \texttt{count\_ops()}, \texttt{depth()}, width, and parameter count from the transpiled (ISA) circuit, partitioning gates into single-qubit, two-qubit, multi-qubit, and arbitrary classes; because the tally is taken after level-3 transpilation, it captures compiler-induced overhead. For CUDA-Q circuits, where no transpiled object is materialised, gate counts are computed analytically from the ansatz structure, via the closed-form QFT expression in experiment~2 and via the per-excitation cost model of Figure~\ref{fig:lcnot-hierarchy} for the LCNot-UCCSD workloads, and validated against the corresponding pre-transpiled Qiskit circuit. This dual accounting facilitates a methodological asymmetry practical to actual application workloads, and not a like-for-like gate-count identity, more concretely it lets compiler overhead on the CPU path and native multi-qubit-gate execution on the GPU path each be described in its own terms. Finally, arithmetic precision is treated as a tunable application-centric variable where, the CUDA-Q GPU path is exercised at both fp32 and fp64 where precision is under test (Sections~\ref{subsubsec:exp1} and~\ref{subsubsec:exp3}), and at fp64 for the other workloads (Sections~\ref{subsubsec:exp2},~\ref{subsubsec:exp4},~\ref{subsubsec:exp5}, and~\ref{subsubsec:exp6}), against the Qiskit CPU default double-precison fp64 statevector reference.
\section{Results and Discussion}
\label{sec:results}

This section reports the empirical outcomes of the evaluation as we move from foundational simulator scaling, where the circuit carries no chemistry and isolates raw state-vector behaviour, through the precision and optimiser characterisation of a minimal molecular workload, into the ansatz- and framework-level efficiency comparison on a molecular test set, and on to the scaling of an application-grade sampling workload toward the 32-qubit frontier. Each outcome constrains the configuration of the next, and the sequence of eliminations (backends, precisions and ans\"atze) converges on a single simulator configuration that is then applied to the nitrogen potential-energy-surface (PES) scan at 32 qubits.

% =====================================================================
\subsection{Foundational Simulator Scaling: GHZ-State Preparation and QFT Mirror Circuits}
\label{sec:results-ghz}

The first experiment fixes the circuit family to GHZ-state preparation and sweeps qubit width against shot count, which removes algorithmic confounders and exposes the bare scaling of each backend up to its compute memory ceiling. The runs were carried out on the local server whose specifications are mentioned in Table~\ref{tab:hardware}. Shot counts of 1{,}000, 10{,}000, 100{,}000 and 1{,}000{,}000 were applied to CUDA-Q targets \texttt{qpp-cpu}, \texttt{nvidia:fp32} and \texttt{nvidia:fp64}, and to the Qiskit \texttt{Aer} and \texttt{StatevectorSampler} (\texttt{Svs}) backends, with each point averaged over ten runs.

\begin{figure}[htbp]
  \centering
  \includegraphics[width=\linewidth]{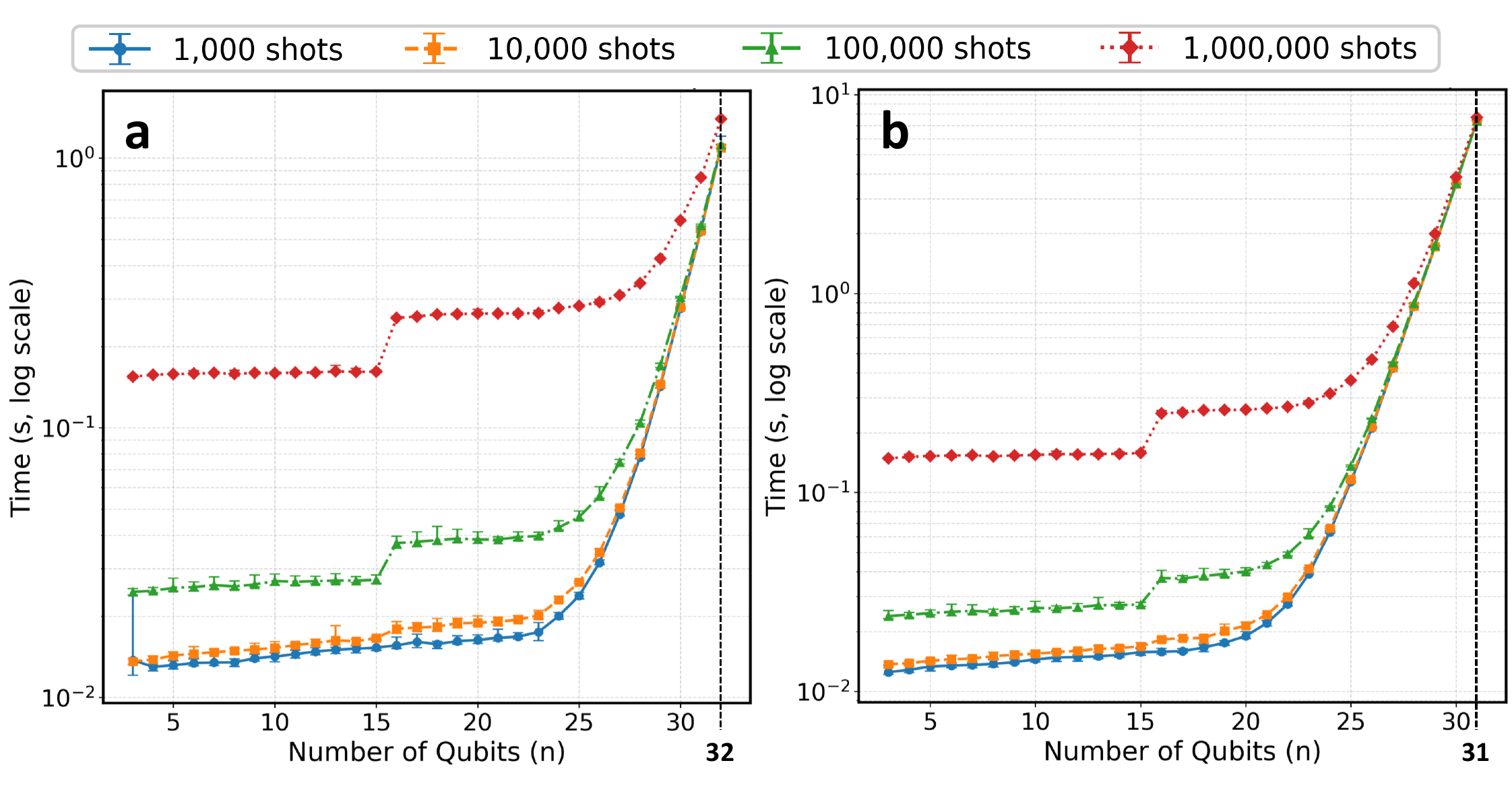}
  \caption{GHZ-state preparation execution time on the CUDA-Q GPU targets. The $x$-axis is the number of qubits and the $y$-axis is the mean execution
    time in seconds (log scale), with error bars denoting $\pm$ min--max over ten runs. Subplots: (a) \texttt{nvidia:fp32} and (b) \texttt{nvidia:fp64}; within each subplot the four curves correspond to shot densities of 1{,}000, 10{,}000, 100{,}000 and 1{,}000{,}000.}
  \label{fig:ghz-gpu}
\end{figure}

\begin{figure}[htbp]
  \centering
  \includegraphics[width=\linewidth]{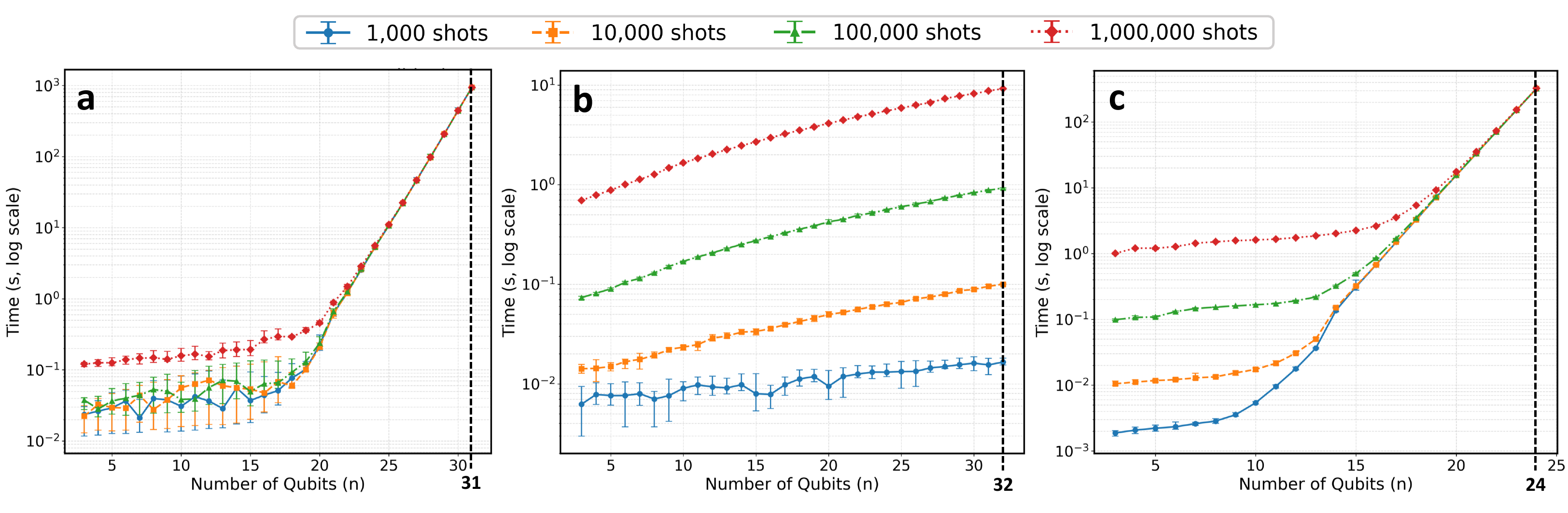}
  \caption{GHZ-state preparation execution time on the CPU backends. The $x$-axis is the number of qubits and the $y$-axis is the mean execution time in seconds (log scale), with error bars denoting $\pm$ min--max over ten runs. Subplots: (a) CUDA-Q \texttt{qpp-cpu}, (b) Qiskit \texttt{Aer} and (c) Qiskit \texttt{StatevectorSampler} (\texttt{Svs}); within each subplot the four curves correspond to shot densities of 1{,}000, 10{,}000, 100{,}000 and 1{,}000{,}000.}
  \label{fig:ghz-cpu}
\end{figure}

\begin{figure}[htbp]
  \centering
  \includegraphics[width=0.85\linewidth]{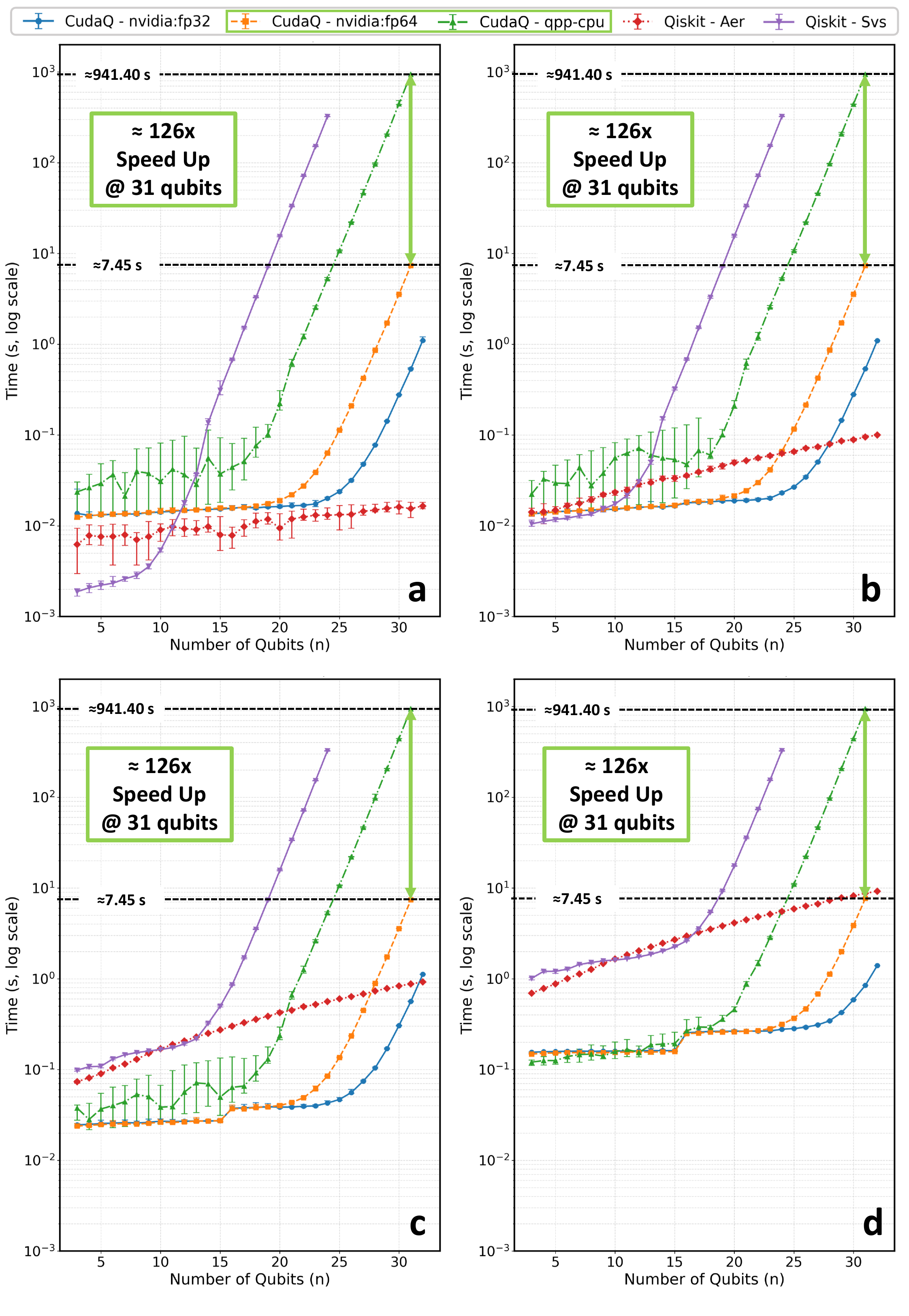}
  \caption{GHZ-state preparation execution time, CPU versus GPU backends overlaid on a common axis. The $x$-axis is the number of qubits and the $y$-axis is the mean execution time in seconds (log scale), with error bars denoting $\pm$ min--max over ten runs. Subplots correspond to shot densities of (a) 1{,}000, (b) 10{,}000, (c) 100{,}000 and (d) 1{,}000{,}000; within each subplot the CUDA-Q GPU targets (\texttt{nvidia:fp32}, \texttt{nvidia:fp64}), the CUDA-Q CPU target (\texttt{qpp-cpu}) and the Qiskit CPU backends (\texttt{Aer}, \texttt{Svs}) are shown together.}
  \label{fig:ghz-cpuvgpu}
\end{figure}

On the GPU (Figure~\ref{fig:ghz-gpu}), the memory envelope sets the hard limit: \texttt{nvidia:fp32} simulates up to 32~qubits and \texttt{nvidia:fp64} up to
31~qubits on the same 48~GB vRAM Nvidia RTX~A6000 Ada GPU, the one-qubit gap being the direct consequence of doubling the bytes stored per amplitude. The runtime profile separates into two regimes. Below about 22~qubits the cost is relatively flat and ordered by shot count, so the work is dominated by shot sampling rather than by state-vector propagation. Above this width, the four shot curves begin to converge and rise steeply, since the exponential growth of the state vector overtakes the fixed shot sampling cost. At the top of the range, double precision is approximately 5.5$\times$ slower than single precision (about 7.67~s versus 1.39~s for the heaviest shot setting), which quantifies the price of the additional precision on top of the one-qubit capacity loss. A reproducible timing artefact at
15--16~qubits is observed consistently across all runs pertaining to a hardware execution artifact and interplay between the CPU, the GPU and the compute memory, whose diagnosis is beyond the scope of this work.

The CPU backends (Figure~\ref{fig:ghz-cpu}) reach 31~qubits (\texttt{qpp-cpu}), 32~qubits (\texttt{Aer}) and 24~qubits (\texttt{Svs}). \texttt{qpp-cpu} and \texttt{Svs} share the same qualitative behaviour: a low, shot-sensitive plateau followed by a sharp state-vector wall, with \texttt{qpp-cpu} showing comparatively high run-to-run variance at small widths. \texttt{Aer} departs from both. Its runtime scales gently and remains separated by shot count across the entire range, so its execution is governed by sampling throughput rather than by a state-vector ceiling for the GHZ experiment, pointing to a different internal simulation strategy. The practical consequence of these distinct profiles is visible in the head-to-head summary: at 24~qubits, \texttt{qpp-cpu} completes in about 5.38~s against about 323.64~s for \texttt{Svs}, a 60$\times$ advantage, while at eight qubits and below (1{,}000 shots) \texttt{Svs} is the fastest of all CPU options. For this GHZ experiment, the \texttt{Aer} CPU backend shows relatively better performance across the width-shot plane.

The CPU-versus-GPU summary (Figure~\ref{fig:ghz-cpuvgpu}) places these results on a common axis. At 31~qubits, the GPU \texttt{nvidia:fp64} is approximately 126$\times$ faster than \texttt{qpp-cpu} (7.45~s versus 941.40~s), and the ratio is essentially independent of shot count. The crossover at which the GPU begins to pay for its fixed launch overhead lies in the 18--22~qubit band; below it the CPU backends, and \texttt{Aer} in particular, even at high shot densities, remain competitive or faster.

\begin{figure}[htbp]
  \centering
  \includegraphics[width=\linewidth]{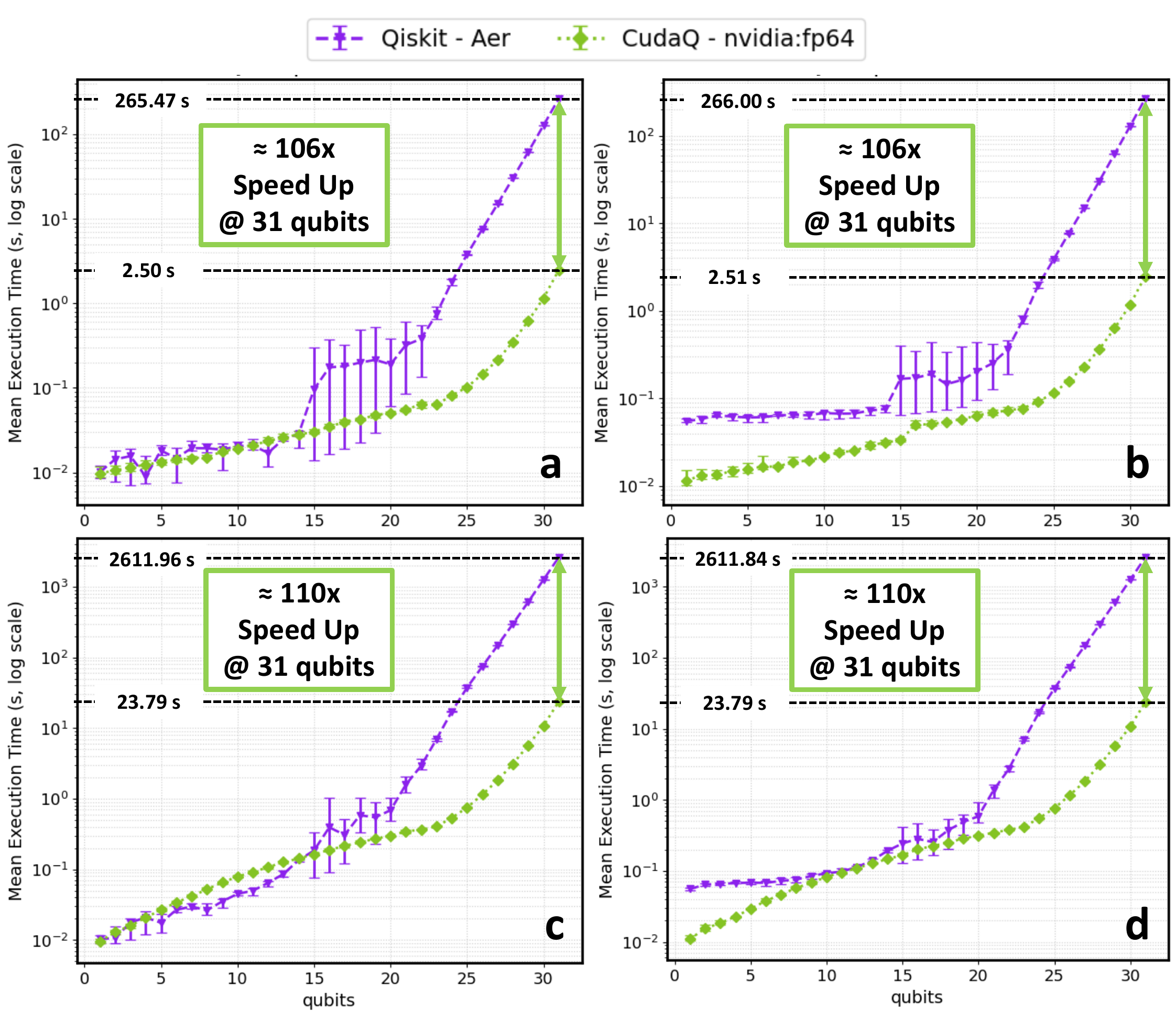}
  \caption{QFT--QFT$^{\dagger}$ mirror-circuit execution time, Qiskit \texttt{Aer} versus CUDA-Q \texttt{nvidia:fp64}. The $x$-axis is the number of qubits and the $y$-axis is the mean execution time in seconds (log scale), with error bars denoting $\pm$ min--max over ten runs. Subplots: (a) one circuit repetition at 10{,}000 shots, (b) one repetition at 100{,}000 shots, (c) ten repetitions at 10{,}000 shots, and (d) ten repetitions at 100{,}000 shots. In each subplot, the dashed guides mark the two backends' mean times at 31 qubits, and the annotation reports the corresponding GPU speed-up.}
  \label{fig:qft-mirror}
\end{figure}

On the basis of these results, we retain one representative backend per device class for the remainder of the study: \texttt{nvidia:fp64} as the GPU target and Qiskit \texttt{Aer} as the CPU target, since it sustained the largest width of any CPU backend (32~qubits) and exhibited the most favourable high-shot scaling. That second observation is also an anomaly: among the CPU backends, \texttt{Aer} alone escapes the state-vector wall that governs \texttt{qpp-cpu} and \texttt{Svs}, and at high shot densities it remains competitive with, and at points ahead of, the GPU. Rather than attribute this to the internals of the \texttt{Aer} implementation, which is outside the scope of this benchmark, we test whether the behaviour is intrinsic to the simulator or specific to the GHZ circuit by repeating the comparison on a structurally different workload. The QFT--QFT$^{\dagger}$ mirror circuit (Figure~\ref{fig:qft-mirror}) serves this purpose: it retains an exact, known output (the input state) while introducing the dense controlled-phase structure absent from GHZ, and the comparison is run for the two retained backends only, \texttt{nvidia:fp64} and \texttt{Aer}.

The mirror-circuit result does not reproduce the GHZ ordering, and the reversal is now quantified. At small widths, the two backends are comparable, but as the qubit count grows, the \texttt{Aer} execution time rises above that of \texttt{nvidia:fp64} and the gap widens monotonically toward the top of the range. At 31 qubits, the GPU leads by roughly two orders of magnitude in every setting: for a single mirror repetition, it completes in 2.50~s against 265.47~s for \texttt{Aer} at 10{,}000 shots and 2.51~s against 266.00~s at 100{,}000 shots, a $\approx$106$\times$ advantage, and for ten mirror repetitions in 23.79~s against 2611.96~s and 2611.84~s at the two shot counts, a $\approx$110$\times$ advantage. Two properties of these numbers are worth recording. First, the top-end times are effectively independent of shot count; the 10{,}000- and 100{,}000-shot runs differ by well under one percent at 31 qubits, which places the workload in the state-vector-bound regime where sampling overhead is negligible, the same regime the GHZ scaling identified at large width. Second, the ten-repetition circuits cost almost exactly ten times the single-repetition circuits, so the measured advantage scales with circuit depth and is a per-run property rather than an artefact of fixed overhead.

The gentle, wall-free scaling that made \texttt{Aer} favourable on GHZ therefore does not carry over to a circuit with dense controlled-phase structure. On the mirror circuit \texttt{Aer} reverts to the behaviour of a conventional CPU state-vector simulator, and the $\approx$106--110$\times$ deficit it concedes to the GPU at 31 qubits is of the same order as the $\approx$126$\times$ gap that \texttt{qpp-cpu} concedes to the GPU on GHZ. We read this as evidence that the GHZ advantage was a property of that circuit family rather than a general simulator characteristic, which is sufficient for the present purpose: \texttt{nvidia:fp64} is confirmed as the GPU representative for the molecular experiments, with \texttt{Aer} retained as the CPU reference whose relative standing is understood to be workload-dependent.

% =====================================================================
\subsection{Precision Floors and Optimiser Behaviour: VQE on H\textsubscript{2}}
\label{sec:results-h2}

The second experiment introduces chemistry at the smallest possible scale, running VQE on H\textsubscript{2} at its equilibrium bond length of 0.7474~\AA{} with a UCCSD ansatz, against the full-configuration-interaction (FCI) reference. The comparison is internal to the NVIDIA stack: CUDA-Q (\texttt{v0.12.0}) against CUDA-QX (\texttt{v0.4.0}), on the same local server, across targets \texttt{qpp-cpu}, \texttt{nvidia:fp32} and \texttt{nvidia:fp64}, and SciPy optimisers L-BFGS-B, SLSQP and COBYLA, with each combination repeated 1{,}000 times. The small-width strips away device-level scaling and isolates the effect of arithmetic precision and optimiser choice on attainable accuracy and cost.

\begin{figure}[htbp]
  \centering
  \includegraphics[width=\linewidth]{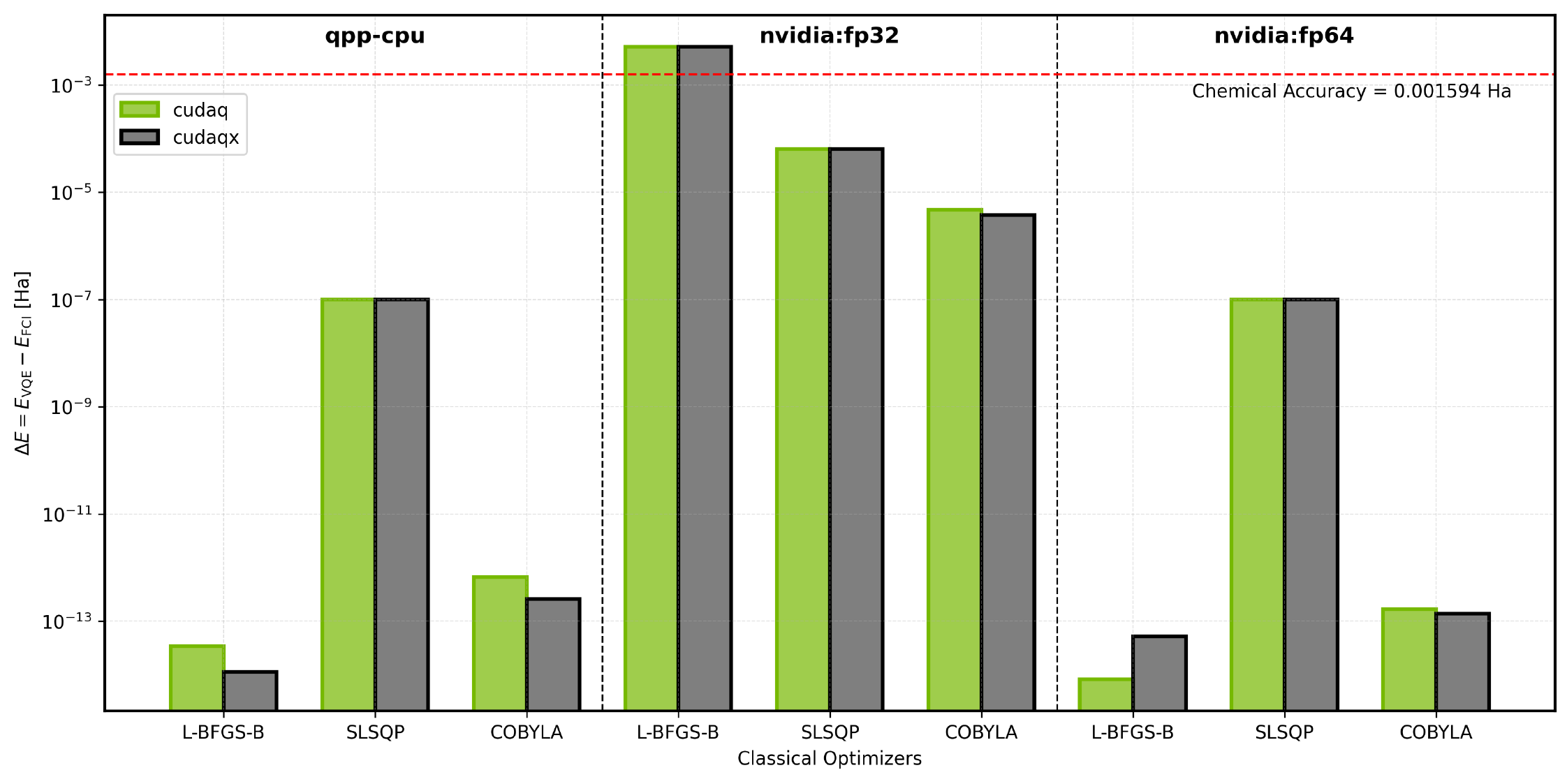}
  \caption{H\textsubscript{2} VQE energy error relative to the FCI reference, $\Delta E = E_{\mathrm{VQE}} - E_{\mathrm{FCI}}$. The $x$-axis groups the three SciPy optimisers (L-BFGS-B, SLSQP, COBYLA) within each of the three targets (\texttt{qpp-cpu}, \texttt{nvidia:fp32}, \texttt{nvidia:fp64}) and the $y$-axis is $|\Delta E|$ in Hartree (log scale); paired bars compare CUDA-Q and CUDA-QX, and the dashed line marks chemical accuracy. Values are from 1{,}000 ideal state-vector runs (zero variance).}
  \label{fig:h2-error}
\end{figure}

\begin{figure}[htbp]
  \centering
  \includegraphics[width=\linewidth]{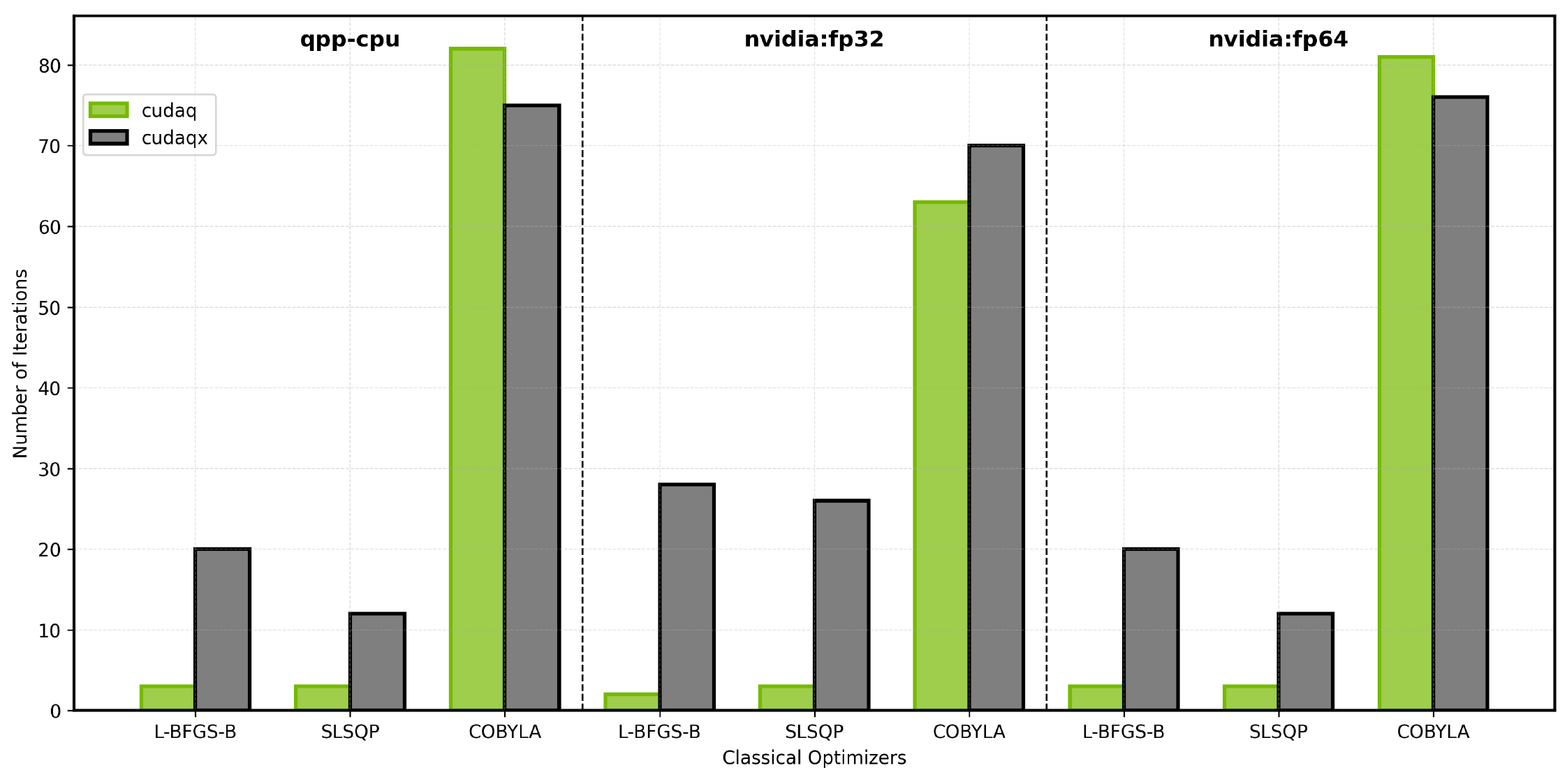}
  \caption{H\textsubscript{2} VQE iterations to convergence. The $x$-axis groups the three optimisers (L-BFGS-B, SLSQP, COBYLA) within each of the three targets (\texttt{qpp-cpu}, \texttt{nvidia:fp32},
    \texttt{nvidia:fp64}) and the $y$-axis is the number of optimiser iterations; paired bars compare CUDA-Q and CUDA-QX.}
  \label{fig:h2-iters}
\end{figure}

\begin{figure}[htbp]
  \centering
  \includegraphics[width=\linewidth]{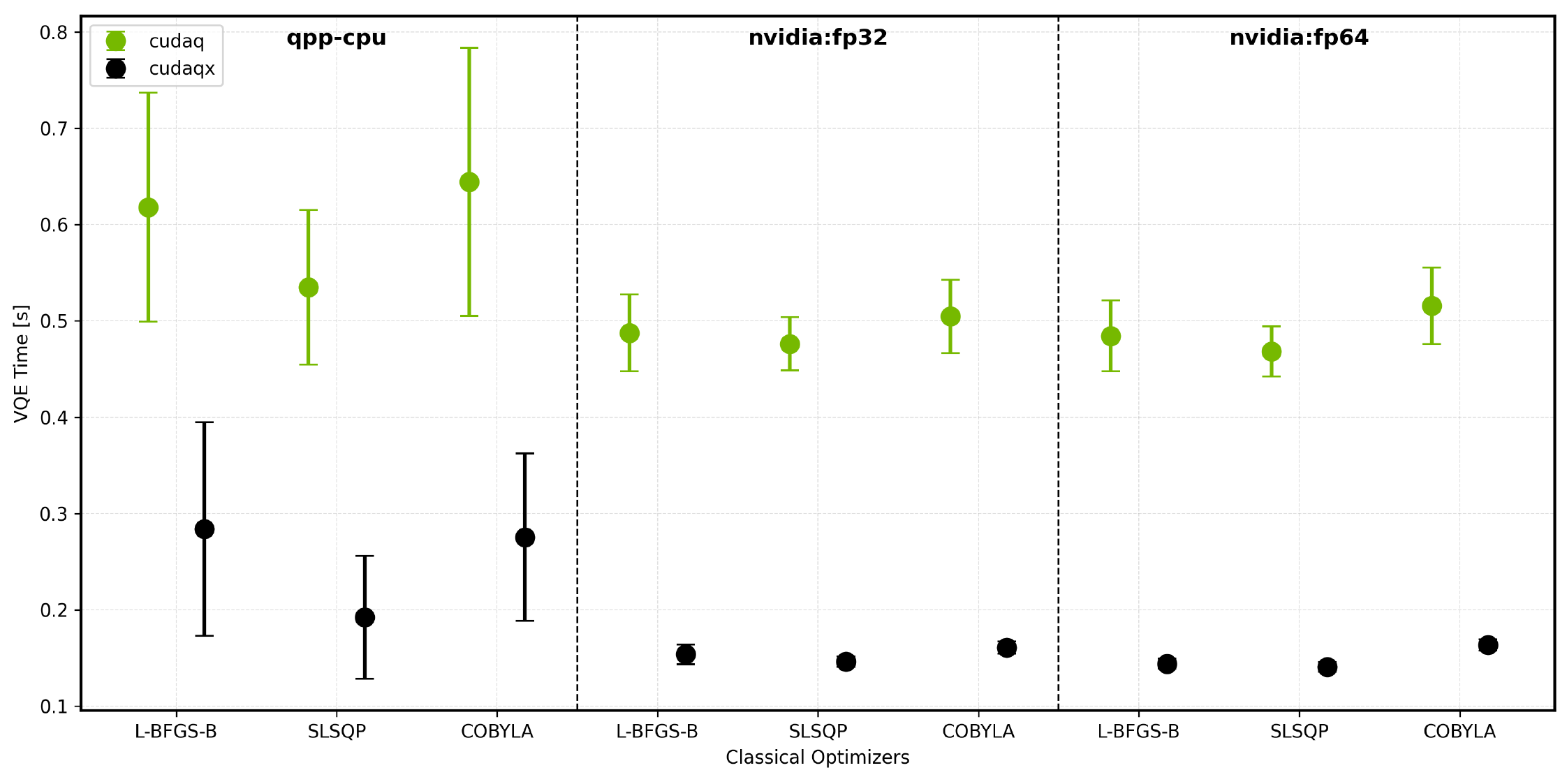}
  \caption{H\textsubscript{2} VQE wall-clock execution time. The $x$-axis groups the three optimisers (L-BFGS-B, SLSQP, COBYLA) within each of the three targets (\texttt{qpp-cpu}, \texttt{nvidia:fp32}, \texttt{nvidia:fp64}) and the $y$-axis is execution time in seconds; points give the mean $\pm$ standard deviation over 1{,}000 executions, comparing CUDA-Q and CUDA-QX.}
  \label{fig:h2-time}
\end{figure}

Accuracy (Figure~\ref{fig:h2-error}) is governed by arithmetic precision rather than by the framework. On \texttt{qpp-cpu} and \texttt{nvidia:fp64}, every optimiser drives the energy error well below the chemical accuracy threshold, with residuals ranging from about $10^{-7}$~Ha (SLSQP) down to $10^{-13}$~Ha (L-BFGS-B and COBYLA). Single precision behaves differently: \texttt{nvidia:fp32} with L-BFGS-B settles at about $3\times10^{-3}$~Ha, above the chemical-accuracy threshold, while its SLSQP and COBYLA runs remain below the line but with markedly larger residuals ($\sim\!7\times10^{-5}$ and $\sim\!5\times10^{-6}$~Ha) than their
double-precision counterparts. CUDA-Q and CUDA-QX produce energy errors that coincide with plotting resolution, so the two frameworks are equivalent in accuracy, and the precision floor is a property of the arithmetic, not of the library. The practical reading is that single precision cannot be relied upon to
reach the chemical accuracy threshold for energy convergence, and that double precision is the safe default for variational chemistry even at this trivial scale.

Optimiser behaviour (Figure~\ref{fig:h2-iters}) is consistent with the cost of the gradient estimate. The gradient-based optimisers L-BFGS-B and SLSQP converge in two to three iterations under CUDA-Q, whereas the gradient-free COBYLA requires between 60 and 82 iterations to reach the same basin. CUDA-QX needs more iterations than CUDA-Q for the gradient-based optimisers (of order 12--28), which, combined with the accuracy equivalence above, preliminarily indicates a difference in the gradient or termination handling between the two libraries rather than a difference in the underlying simulation subject to further study.

Timing (Figure~\ref{fig:h2-time}) inverts the intuition carried over from the GHZ scaling. At this width, the device contributes little: GPU and CPU execution times are of the same order, and single and double precision are indistinguishable, confirming that the fp64 cost seen at the top of the GHZ range does not manifest at four qubits. The separation that does appear is between frameworks. CUDA-QX is an application layer built on CUDA-Q, and both ultimately dispatch to the same \texttt{cuStateVec} state-vector engine, so the energies are equivalent; the differences are confined to the execution path around that engine. Relative to CUDA-Q, CUDA-QX runs with lower and more tightly clustered per-run times, but converges in more optimiser iterations at default settings, so its advantage in wall-clock time is accompanied by a cost in convergence
economy.

Two selections follow from this experiment and are carried through the remaining molecular work. First, CUDA-Q is adopted as the common implementation: it is the lower-level framework on which CUDA-QX is built, and standardising on it avoids the default iteration inflation observed for CUDA-QX while leaving the underlying
accuracy unchanged. Second, \texttt{nvidia:fp64} is adopted over \texttt{nvidia:fp32}: the accuracy results show that single precision may not be relied upon to reach the chemical accuracy threshold for energy convergence. At the small-width limit, then, framework and optimiser selection govern both cost and convergence; the device of choice (CPU/GPU) seems to be immaterial, and precision is one of the dominant levers that decides whether the chemical accuracy threshold is reachable.

% =====================================================================
\subsection{Ansatz and Framework Efficiency: VQE with UCCSD versus LCNot-UCCSD}
\label{sec:results-uccsd}

The third experiment moves from a single molecule to a five-molecule set and introduces the LCNot-UCCSD ansatz as an alternative to standard UCCSD, benchmarking both ansatze across both frameworks (Qiskit and CUDA-Q) under matched geometries and active spaces. The basis set chosen is STO-3G, and the optimiser is
SLSQP with a tolerance of $10^{-6}$, and each combination is repeated ten times. Every molecule except H\textsubscript{2} is treated under the frozen-core approximation, so the reported qubit widths refer to the correlated active space. Energies are taken as ideal state-vector expectation values, so no shot noise persists. Table~\ref{tab:mol-ref} lists the qubit-Hamiltonian term counts, parameter counts, and CCSD and FCI reference energies that anchor the accuracy analysis.

\begin{table}[htbp]
  \centering
  \caption{Qubit-Hamiltonian terms, variational parameter counts, and PySCF reference energies for the molecular test set. CCSD energies use the frozen-core approximation. Term counts marked with $^{*}$/$^{+}$ are the Qiskit values; CUDA-Q counts differ slightly (BeH\textsubscript{2}: 1003 for LCNot-UCCSD, 1009 for UCCSD; H\textsubscript{2}O: 1055 for LCNot-UCCSD, 1059 for UCCSD).}
  \label{tab:mol-ref}
  \begin{tabular}{lrrrr}
    \toprule
    Molecule & Hamiltonian terms & Parameters & $E_{\mathrm{CCSD}}$ [Ha] & $E_{\mathrm{FCI}}$ [Ha] \\
    \midrule
    H\textsubscript{2}   & 15           & 3  & $-1.1278426051$  & $-1.1278425043$ \\
    LiH                  & 276          & 24 & $-7.8809467879$  & $-7.8811783517$ \\
    HF                   & 276          & 24 & $-98.6000482842$ & $-98.6000687955$ \\
    BeH\textsubscript{2} & $1007^{*}$   & 92 & $-15.5470835721$ & $-15.5477979299$ \\
    H\textsubscript{2}O  & $1063^{+}$   & 92 & $-75.0153353850$ & $-75.0155335184$ \\
    \bottomrule
  \end{tabular}
\end{table}

\begin{figure}[htbp]
  \centering
  \includegraphics[width=\linewidth]{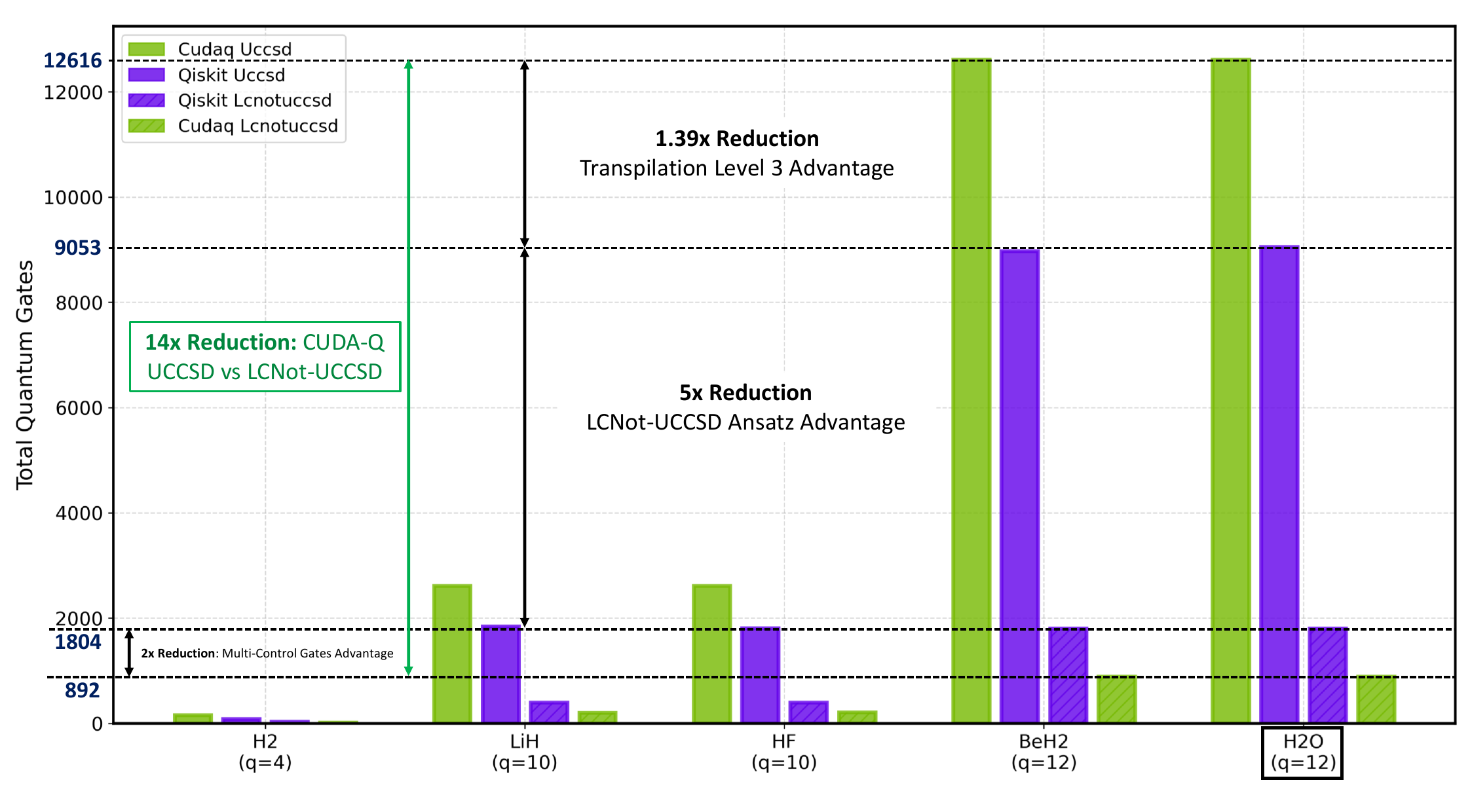}
  \caption{Quantum resource estimation (total gate count) for the UCCSD and LCNot-UCCSD ans\"atze under Qiskit and CUDA-Q. The $x$-axis lists the molecular test set (H\textsubscript{2}, LiH, HF, BeH\textsubscript{2}, H\textsubscript{2}O, with qubit widths) and the $y$-axis is the total gate
    count; the four grouped bars per molecule are CUDA-Q UCCSD, Qiskit UCCSD, Qiskit LCNot-UCCSD and CUDA-Q LCNot-UCCSD (bars: mean $\pm$ min--max over ten runs).}
  \label{fig:uccsd-gates}
\end{figure}

\begin{figure}[htbp]
  \centering
  \includegraphics[width=\linewidth]{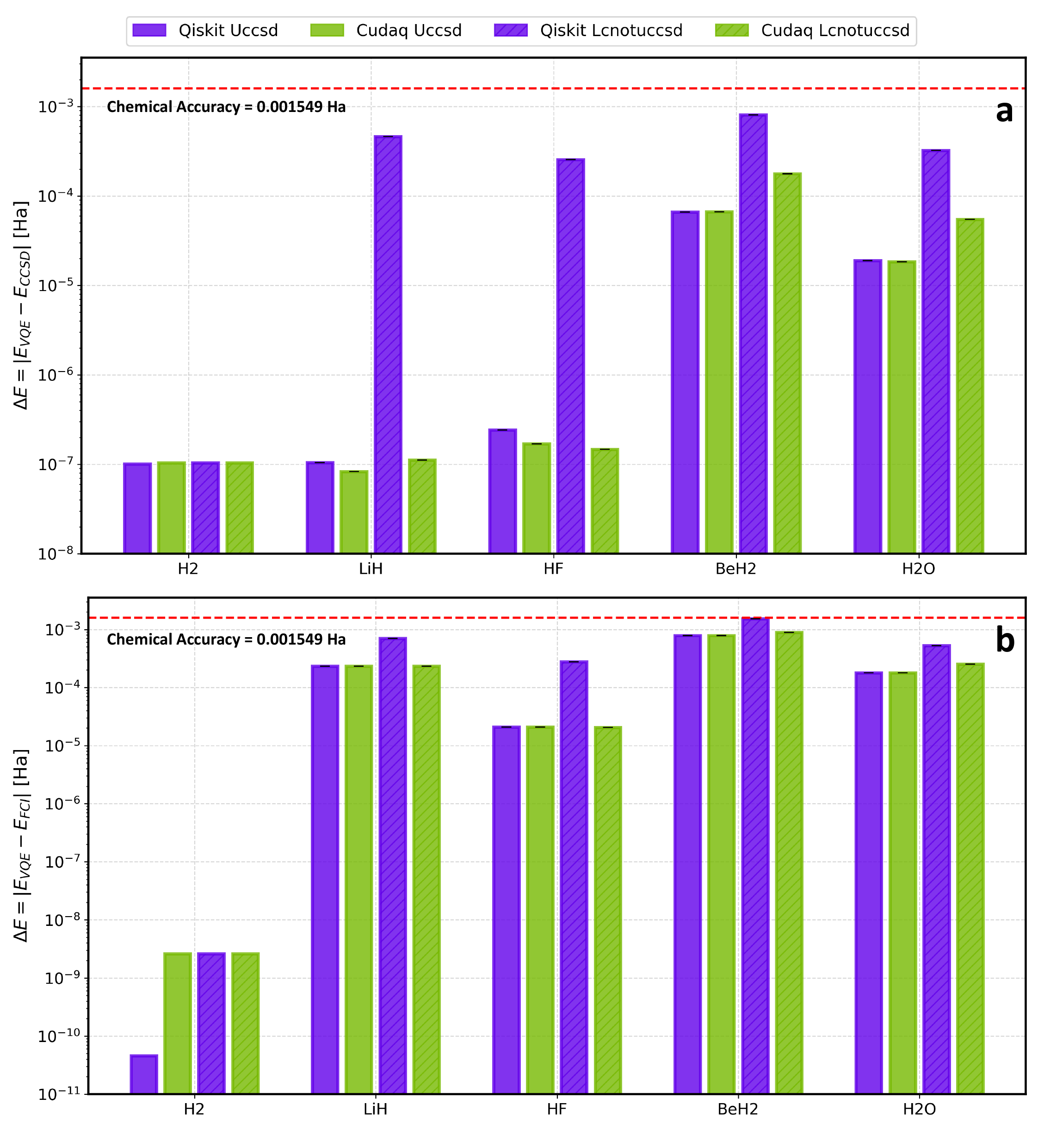}
  \caption{VQE energy error for the UCCSD and LCNot-UCCSD ans\"atze under both frameworks. The $x$-axis lists the molecular test set and the $y$-axis is $|\Delta E|$ in Hartree (log scale), with the dashed line marking chemical accuracy; the four grouped bars per molecule are Qiskit UCCSD, CUDA-Q UCCSD, Qiskit LCNot-UCCSD and CUDA-Q LCNot-UCCSD. Subplots: (a) error  against the CCSD reference, $|E_{\mathrm{VQE}} - E_{\mathrm{CCSD}}|$, and (b) error against the FCI reference, $|E_{\mathrm{VQE}} - E_{\mathrm{FCI}}|$.}
  \label{fig:uccsd-acc}
\end{figure}

\begin{figure}[htbp]
  \centering
  \includegraphics[width=\linewidth]{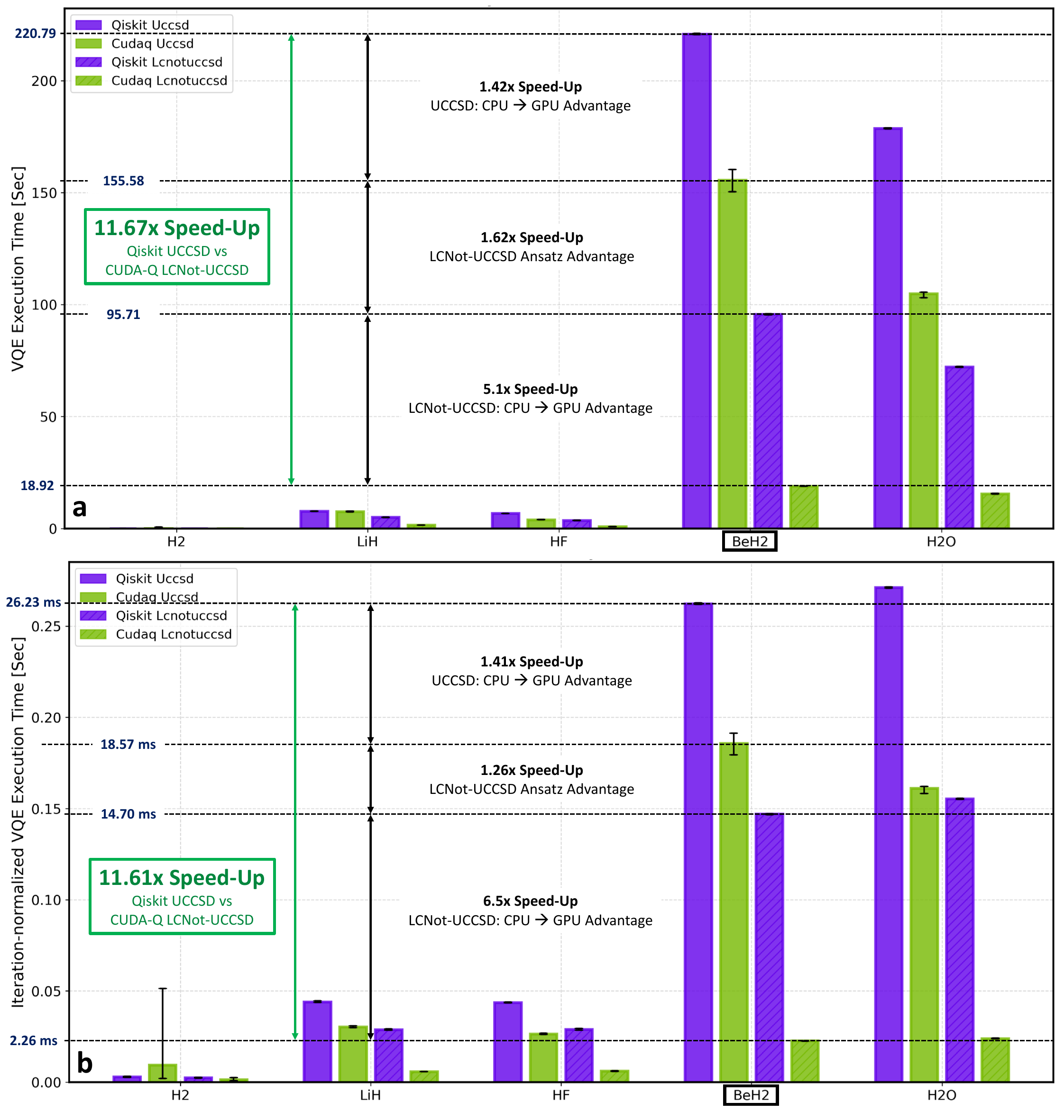}
  \caption{VQE execution time for BeH\textsubscript{2} (12~qubits, 4~electrons), the molecule with the highest iteration count in the set. The bars are Qiskit UCCSD, CUDA-Q UCCSD, Qiskit LCNot-UCCSD and CUDA-Q LCNot-UCCSD (mean $\pm$ min--max over ten runs). Subplots: (a) total execution time in seconds, and (b) execution time normalised by the number of optimiser iterations, in milliseconds.}
  \label{fig:uccsd-time}
\end{figure}

\begin{figure}[htbp]
  \centering
  \includegraphics[width=\linewidth]{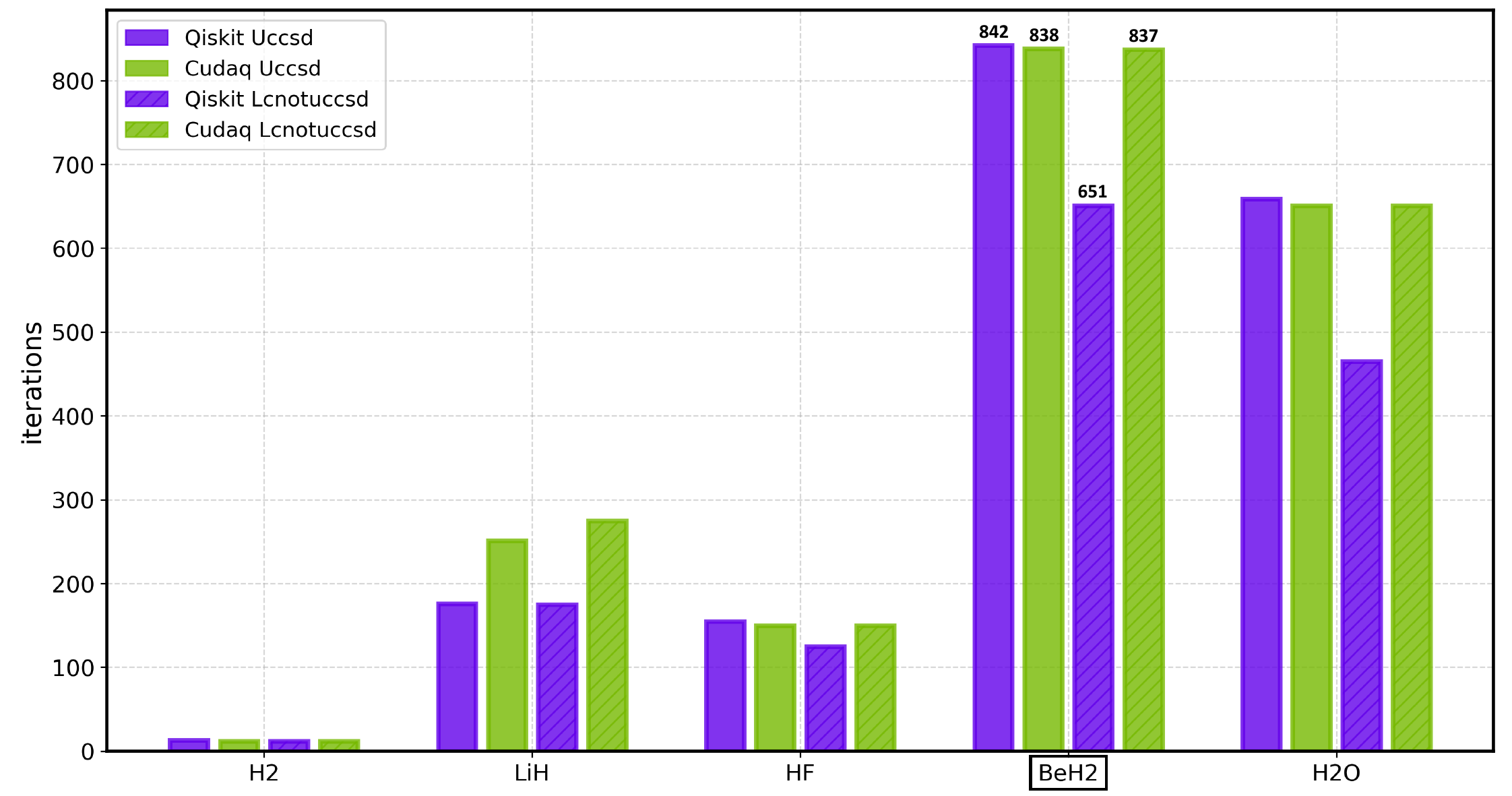}
  \caption{VQE iterations to convergence for the molecular test set. The $x$-axis lists the molecules and the $y$-axis is the number of optimiser iterations; the four grouped bars per molecule are Qiskit UCCSD, CUDA-Q UCCSD, Qiskit LCNot-UCCSD and CUDA-Q LCNot-UCCSD (bars: mean $\pm$ min--max over ten runs).}
  \label{fig:iterations-vqe}
\end{figure}

Resource estimation (Figure~\ref{fig:uccsd-gates}) establishes the necessary baseline for this experiment, where we take H\textsubscript{2}O at 12~qubits as the reference case, since it is the largest molecule in the set and exposes the full reduction chain; the gate count falls from 12616 for CUDA-Q UCCSD to 892 for CUDA-Q LCNot-UCCSD, a 14$\times$ reduction. This total decomposes into three independent contributions: Qiskit transpilation level~3 trims the UCCSD circuit by 1.39$\times$ relative to the CUDA-Q UCCSD count, the LCNot-UCCSD ansatz itself removes a further 5$\times$, and CUDA-Q's native multi-control gates contribute a final 2$\times$ over the Qiskit LCNot-UCCSD circuit. Based on this, we may infer that the ansatz is the largest lever, followed by the framework's native gate set, which adds a smaller reduction on top of it. The same hierarchy holds across LiH, HF, BeH\textsubscript{2} and H\textsubscript{2}O, with the absolute gap widening as molecule size grows across the five molecules under study.

Accuracy (Figure~\ref{fig:uccsd-acc}) confirms that the gate savings are not purchased at the cost of correctness. Measured against CCSD, every framework and ansatz combination remains within chemical accuracy for all five molecules. CUDA-Q LCNot-UCCSD tracks full UCCSD closely, while Qiskit LCNot-UCCSD carries consistently larger residuals, most visibly for LiH and HF, where its error is two to three orders of magnitude above the UCCSD baseline, and against FCI it approaches the chemical-accuracy threshold for BeH\textsubscript{2}. The difference between the two LCNot-UCCSD implementations reflects their respective standard circuit transpilation handling rather than the ansatz itself and warrants further study deferred to future work.

The timing decomposition (Figure~\ref{fig:uccsd-time}a) ties the gate and device effects together. BeH\textsubscript{2} carries the highest iteration count of any molecule in the set across all four framework, ansatz combinations, so it is
used as the timing exemplar (Figure~\ref{fig:iterations-vqe}). Its execution time falls from 220.79~s (Qiskit UCCSD) to 18.92~s (CUDA-Q LCNot-UCCSD), an 11.67$\times$ end-to-end speed-up that factorises into 1.42$\times$ from moving UCCSD onto the GPU, 1.62$\times$ from switching to the LCNot-UCCSD ansatz, and 5.1$\times$ from running that ansatz on the GPU. The iteration counts behind these times are broadly comparable across configurations: 842, 838, 651 and 837 for the four BeH\textsubscript{2} cases, and the Qiskit LCNot-UCCSD variant occasionally converges in fewer iterations despite its lower accuracy. The speed-up is
therefore driven by per-iteration cost, not by faster convergence, which the normalised metric (Figure~\ref{fig:uccsd-time}b)isolates: 26.23~ms, 18.57~ms, 14.70~ms and 2.26~ms per iteration along the same chain, an 11.61$\times$ reduction that mirrors the wall-clock figure
almost exactly. The compounding of an ansatz-level gate reduction by moving from UCCSD to the LCNot-UCCSD variant along with GPU acceleration is the central efficiency result carried forward into the application-width experiments.

% =====================================================================
\subsection{Scaling to Application Width: Quantum Sampling on Hydrogen Chains}
\label{sec:results-sampling}

The fourth experiment pushes a single workload, LCNot-UCCSD sampling, across a family of hydrogen chains of increasing length, from H\textsubscript{2} to H\textsubscript{16}, spanning 4 to 32 qubits in circuit width. Hydrogen chains are chosen because, for a given qubit count, they carry the largest number of single and double excitations, and therefore generate the deepest, gate-heavy circuits available in the molecular series, which makes them the most demanding stress test of the two stacks at the 32-qubit frontier the N$_2$ PES requires. Each chain uses a bond distance of 1.0~\AA{} and 10{,}000 shots. The two stacks differ deliberately in how they prepare the circuit as detailed in Section~\ref{subsubsec:exp5}.

\begin{figure}[htbp]
  \centering
  \includegraphics[width=\linewidth]{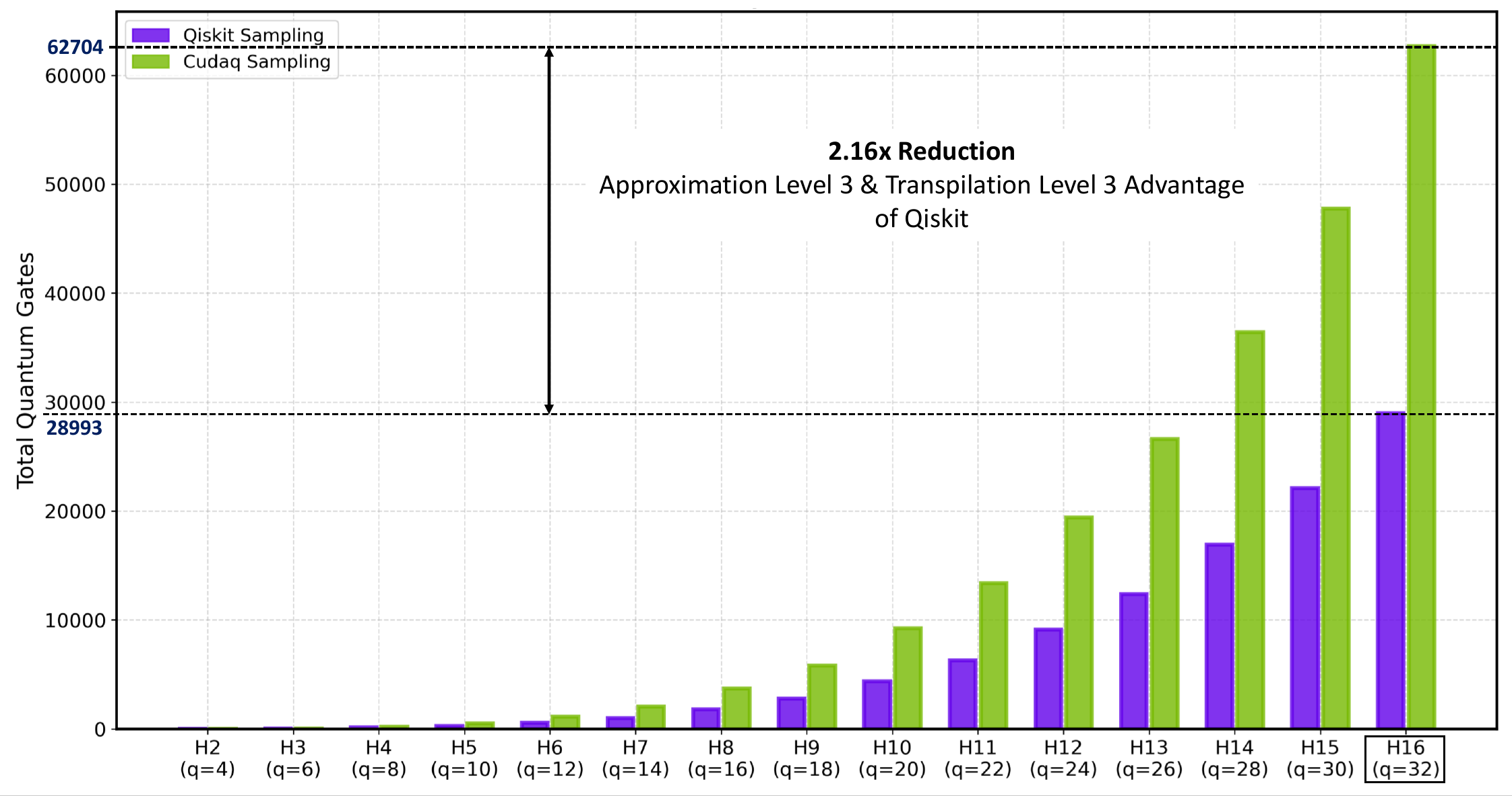}
  \caption{Quantum resource estimation for LCNot-UCCSD sampling on the hydrogen chains. The $x$-axis lists the chains
    H\textsubscript{2}--H\textsubscript{16} (4--32 qubits) and the $y$-axis is the total gate count; the two bars per chain are Qiskit sampling and CUDA-Q sampling (bars: mean $\pm$ min--max over ten runs).}
  \label{fig:sampling-gates}
\end{figure}

\begin{figure}[htbp]
  \centering
  \includegraphics[width=\linewidth]{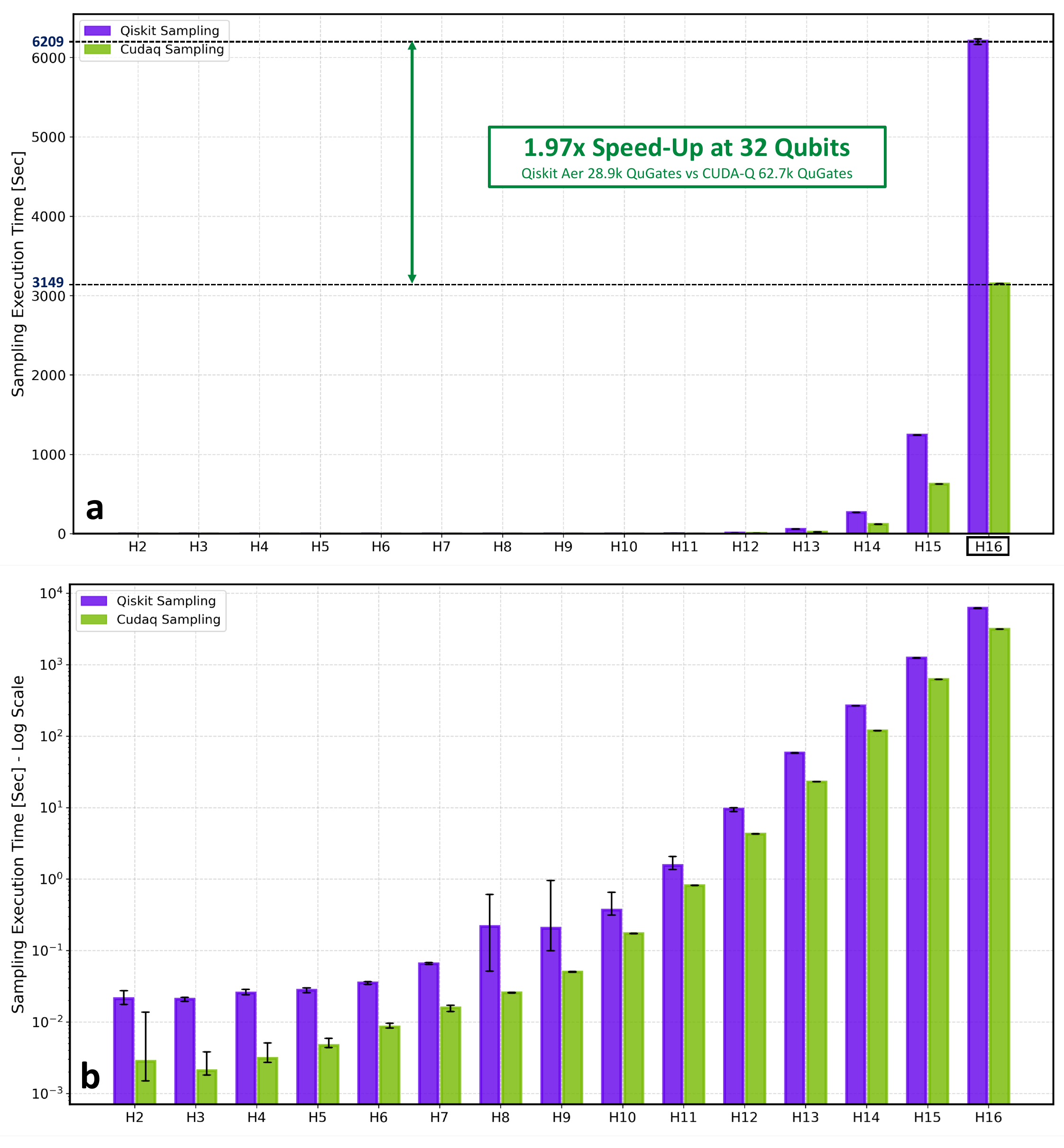}
  \caption{Quantum-circuit sampling execution time on the hydrogen chains, Qiskit versus CUDA-Q. The $x$-axis lists the chains H\textsubscript{2}--H\textsubscript{16} (4--32 qubits), and the $y$-axis is the sampling execution time in seconds (bars: mean $\pm$ min--max over ten runs). Subplots: (a) linear scale and (b) log scale.}
  \label{fig:sampling-time}
\end{figure}

The resource estimate (Figure~\ref{fig:sampling-gates}) reverses the gate-count ordering seen in the previous experiments. Here Qiskit produces the smaller circuit at 28993 gates against 62704 for CUDA-Q at 32~qubits, a 2.16$\times$ reduction, an artefact of its approximation- and transpilation-level settings compressing the LCNot-UCCSD decomposition compared to that of CUDA-Q which executes unoptimised. It is to be noted that the earlier 2$\times$ advantage of CUDA-Q's native multi-control gates does not appear under these settings, which makes explicit that gate-count outcomes are a property of the compilation pipeline as configured (Section~\ref{subsubsec:exp5}) and not a fixed property of either framework.

The timing result (Figure~\ref{fig:sampling-time}) is more consequential where CUDA-Q is faster than Qiskit at every chain length, and at 32~qubits it completes sampling in 3149.7~s against 6209.3~s for Qiskit, a 1.97$\times$ advantage. This speed-up is obtained even though CUDA-Q runs 2.16$\times$ more gates, so the GPU's state-vector throughput more than compensates for the larger circuit. The margin is not constant: it is close to an order of magnitude at the small widths, where absolute times are sub-second, and the GPU's high relative variance reflects launch and measurement overhead, and it narrows to about 2$\times$ at 32~qubits as the unoptimised gate count erodes the advantage. Below roughly 26~qubits, the absolute times are short enough that the device choice is immaterial; the exponential wall, and with it the GPU's value, appears only at the largest widths. This establishes that 32~qubits is a workable operating point for the GPU and at that width, a net 2$\times$ wall-clock advantage is available even against a more aggressively compiled CPU baseline.

% =====================================================================
\subsection{N\textsubscript{2} Potential-Energy-Surface Scan via QSCI(LCNot-UCCSD)--RBM}
\label{sec:results-n2}

The preceding experiments narrow the configuration space by successive elimination across CPU and GPU backends, single and double precision, and the UCCSD and LCNot-UCCSD ans\"atze to converge on CUDA-Q \texttt{nvidia:fp64} as the state-vector simulator and LCNot-UCCSD as the ansatz for the nitrogen PES scan presented here. The scan covers twenty bond lengths spanning the compressed ($r < 0.99$~\AA), near-equilibrium and dissociation ($r > 1.32$~\AA) regions in the cc-pVDZ basis, and each geometry is evaluated once, so the curves carry no error bars. The simulation is an ideal, noiseless state-vector evaluation, which stands in for a representative fault-tolerant execution of the circuit with device noise abstracted away. Two methods are compared at every geometry, both built on the LCNot-UCCSD ansatz, including SQD, and QSCI-RBM. Hartree--Fock, MP2, CCSD and CASCI$(10e, 16o)$ curves are included as references, with CASCI$(10e, 16o)$ serving as the active-space exact benchmark against which the two quantum methods are evaluated.

\begin{figure}[htbp]
  \centering
  \includegraphics[width=\linewidth]{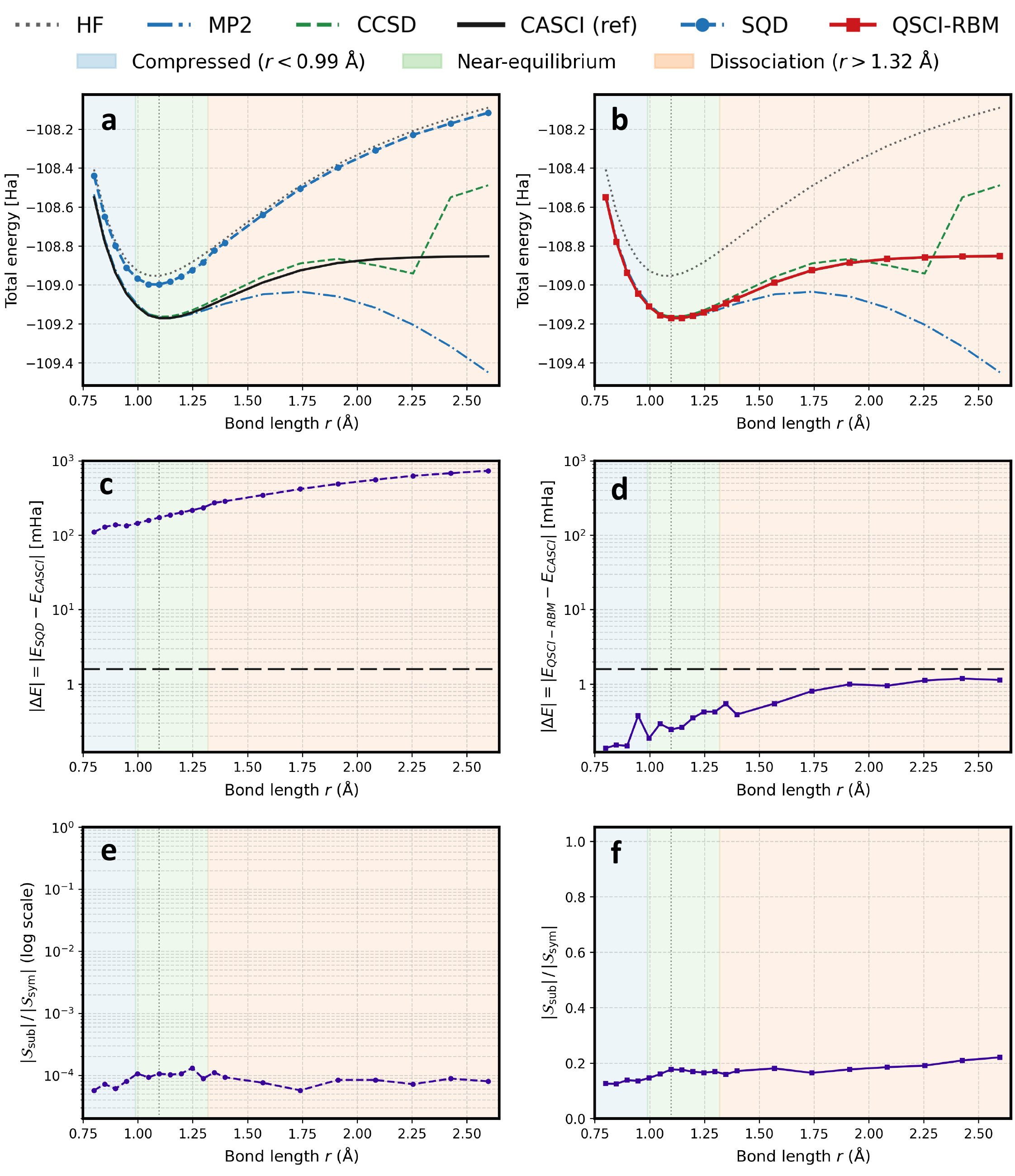}
  \caption{N\textsubscript{2} $(10e, 16o)$ potential-energy-surface scan in the cc-pVDZ basis. In every panel the $x$-axis is the bond length $r$ in \AA{}ngstr\"om, and shaded bands mark the compressed ($r < 0.99$~\AA), near-equilibrium and dissociation ($r > 1.32$~\AA) regions. Subplots: (a) total energy (Ha) versus $r$ for HF, MP2, CCSD, CASCI$(10e,16o)$ and SQD; (b) the same references with QSCI-RBM; (c) absolute energy deviation $|E_{\mathrm{SQD}} - E_{\mathrm{CASCI}}|$ in mHa (log scale), with the dashed line at the $1.59$~mHa chemical-accuracy threshold; (d) the corresponding deviation $|E_{\text{QSCI-RBM}} - E_{\mathrm{CASCI}}|$; (e) the diagonalization subspace coverage ratio $\eta_{\mathrm{sub}} = |\mathbb{S}_{\mathrm{sub}}|/|\mathbb{S}_{\mathrm{sym}}|$ for SQD (log scale); and (f) the same ratio for QSCI-RBM (linear scale).}
    \label{fig:n2-pes}
\end{figure}

Both methods trace a smooth PES across the full scan (Figure~\ref{fig:n2-pes}a,b), with no discontinuities through the dissociation
region where the triple bond breaks and static correlation is strongest. Whereas the two diverge sharply in accuracy, where SQD sits essentially close to the Hartree--Fock curve, and its deviation from the CASCI reference is correspondingly large: across the surface it remains orders of magnitude above the $1.59$~mHa chemical-accuracy threshold (Figure~\ref{fig:n2-pes}c), so within this active
space SQD recovers little correlation beyond the mean-field solution. QSCI-RBM, by contrast, stays within the chemical accuracy threshold of the CASCI$(10e, 16o)$ reference across the entire scan (Figure~\ref{fig:n2-pes}d), including through the dissociation region.

QSCI-RBM attains the chemical accuracy threshold across the scan while its classical diagonalization samples a fraction of the symmetry-adapted configuration space, on the order of 15--20\%, with a mean near $0.17$ and a maximum close to $0.22$ across the scan (Figure~\ref{fig:n2-pes}f). Here, the RBM-guided selection places its truncated subspace on the configurations that carry the correlation. The coverage QSCI-RBM uses may be considered as an upper bound rather than a requirement, as per-geometry hyperparameter optimisation of the selection would be expected to reduce the diagonalization subspace further at fixed accuracy, warranting further study. Nevertheless, from a computational perspective, the result establishes that the $(10e, 16o)$ surface, out of reach of the earlier 16-qubit CPU workflow, is computed on a single GPU in double precision.

% =====================================================================
\subsection{Cross-Experiment Synthesis and Selection Guidance}
\label{sec:results-synthesis}

Read together, the experiments presented in this work yield a configuration guidance rather than a single ranking. The device (CPU/GPU) crossover appears to be consistent where GPU acceleration pays for its fixed overhead only above roughly 18--26 qubits, depending on workload, and below that width CPU backends remain competitive or faster, with Qiskit \texttt{Aer} particularly strong at high shot densities and \texttt{Svs} fastest at the smallest widths and low shot counts. That competitiveness is, however, workload-dependent, where the \texttt{Aer} advantage seen on GHZ is not replicable in the QFT--QFT$^{\dagger}$ mirror circuits, where the GPU is faster at application-relevant widths, indicating that the backend rankings established on one circuit family should not be assumed to transfer to another. In cases where the GPU shows a time-to-solution advantage, the measured advantage spans a wide range, about 2$\times$ for the gate-heavy 32-qubit sampling workload, 5$\times$ for LCNot-UCCSD VQE, and in the order of 100$\times$ for GHZ state-vector preparation, and the QFT--QFT$^{\dagger}$ mirror circuit at 31 qubits. This indicates that the speed-up is a function of how state-vector-bound the workload is, and not a fixed multiplier. We may also assert that the arithmetic precision is non-negotiable, where single precision both increases the qubit ceiling by one but more importantly, fails to reach the chemical accuracy threshold for energy convergence, so double precision is the default for variational and sampling quantum simulation workloads. Circuit-level economy compounds with device choice where the LCNot-UCCSD ansatz reduces gate counts by up to 14$\times$ relative to UCCSD, and stacking that reduction with GPU execution produces the largest end-to-end gains observed in the molecular experiments. Applied together in the N$_2$ PES scan, these selections place the N\textsubscript{2} $(10e, 16o)$ surface, a 32-qubit problem, as the final application demonstration of this work using GPU acceleration.

\section{Conclusion}
\label{sec:conclusion}

In this work, we set out to answer a practitioner's question: given a fixed application objective and a finite hardware budget, which simulation stack, precision, and ansatz should carry the computation, and where does that choice stop mattering? We approach it as a single chain of six experiments in which each result narrows the space of sensible configurations for the next. The foundational benchmarks fixed the engine and precision, the quantum chemistry algorithm benchmarks fixed the ansatz, and the capstone demonstration spent the accumulated choices on a problem that a prior QSCI(LCNot-UCCSD)-RBM workflow could not reach.

The foundational tier of experiments establishes two governing points where the H\textsubscript{2} VQE comparison retained CUDA-Q over CUDA-QX as the GPU path, and the GHZ scaling study exposes the underlying cost structure: double precision (\texttt{nvidia:fp64}) is the precision-safe target for the chemistry that follows, but it forfeits one qubit of state-vector capacity relative to single precision under fixed memory, with memory rather than arithmetic setting the ceiling on simulable width. That work also surfaces an anomaly where Qiskit \texttt{Aer} alone among the CPU backends escapes the state-vector wall that constrained \texttt{qpp-cpu} and \texttt{Svs}, sustaining the largest CPU width, and at high shot densities remained competitive with, and at points ahead of, the GPU. In order to assess whether this behaviour was intrinsic or an artefact of the GHZ family, we introduce the QFT--QFT\textsuperscript{$\dagger$} mirror circuits test, which did not reproduce the GHZ ordering, as the GPU state-vector simulator was faster at application-relevant widths on a workload with dense controlled-phase structure.

The following application tier experiments first fix the ansatz where across the five-molecule VQE set, LCNot-UCCSD reproduces the accuracy of standard UCCSD while reducing the entangling-gate footprint, and the hydrogen-chain sampling study extends the comparison to the \num{32}-qubit boundary (H\textsubscript{16}) under a framework-optimal throughput protocol in which each stack is timed at its own best-case operating point. Combined with the foundational results, this converges on a single justified configuration, CUDA-Q \texttt{nvidia:fp64} with LCNot-UCCSD, as the endpoint of a constructive elimination. That configuration is then applied to the QSCI(LCNot-UCCSD)--RBM workflow to complete a \num{20}-point N\textsubscript{2} potential-energy-surface scan in a cc-pVDZ \num{16}-orbital / \num{10}-electron active space (\num{32} spin orbitals, \num{32} qubits), doubling the active-space width of the prior CPU-bound study~\cite{KSV2026FX700} through GPU acceleration of both the state-vector simulation and the PyTorch-based RBM. The QSCI-RBM surface remained within chemical accuracy threshold of the CASCI reference across the full scan while diagonalizing about \SIrange{15}{20}{\percent} of the symmetry-adapted configuration space, whereas the SQD baseline tracked the Hartree--Fock surface with deviations orders of magnitude larger.

Stated compactly, GPU-accelerated double precision delivers consistent speedups, typically $2\times$--$10\times$ for tractable application workloads and up to $100\times$ at the widest circuits, and governs the maximum simulable system size under fixed compute memory, while CPU state-vector backends and Qiskit \texttt{Aer} remain competitive, and occasionally preferable, at low qubit counts and high shot densities. We would also like to note that the boundaries of this work also mark the natural extensions where all results are obtained under noiseless state-vector simulation and characterise the classical simulators and the algorithmic workflow rather than physical-device behaviour, introducing device noise models and executing the reduced-depth LCNot-UCCSD circuits on real backends with native multi-qubit controlled gates support and error mitigation is the next step in testing whether the ansatz-level gate savings survive under noise. The \num{32}-qubit active space is the ceiling of single-GPU double-precision simulation for the chosen GPUs, not of the workflow itself, and distributed or multi-GPU execution, tensor-network backends where circuit structure permits, and mixed-precision strategies that reclaim the qubit of width forfeited by \texttt{nvidia:fp64} will all push the accessible active space further. Finally, the RBM hyperparameters are held fixed across the scan even though the subspace fraction required for chemical accuracy may vary along the surface, so adaptive per-geometry schedules may tighten the coverage figures at the stretched geometries where correlation is strongest. We regard the six-experiment methodology as a more portable contribution, applicable wherever a finite compute budget must be matched to a specific application objective.

\section*{Data Availability}
The datasets generated and analysed during the current study are available in the Zenodo repository (DOI: 10.5281/zenodo.21189705).

\section*{Acknowledgements}
The authors gratefully acknowledge the I-HUB Quantum Technology Foundation (I-HUB QTF) and NVIDIA for providing access to high-performance GPU-enabled infrastructure through the 2025 "Join the Challenge to be Quantum Ready" initiative. Specifically, the GPU-accelerated CUDAQ evaluations for the Variational Quantum Eigensolver (VQE) benchmarks of the five different molecules reported in this work were executed on the provided DGX A100 platform. The authors extend their sincere appreciation to the advisors of Qclairvoyance Quantum Labs for their institutional support, constructive technical discussions, and continuous inspiration throughout the formulation and preparation of this work.

 \section*{Funding}
No funding was received for this research.

\section*{Competing Interests}
R.M. is a paid consultant at Qclairvoyance Quantum Labs. The other authors declare no competing interests.

\bibliography{references}

\end{document}